%% file: main.tex
\documentclass[creativecommons]{eptcs}

\usepackage{custom-commands}

\usepackage{tikz-mlsl}

\usepackage{custom-document-commands}

\providecommand{\event}{FVAV 2017}

\title{Monitoring of Traffic Manoeuvres with \mbox{Imprecise Information}\thanks{Work of the author is supported by the Deutsche Forschungsgemeinschaft (DFG) within the Research Training Group DFG GRK 1765 SCARE.}}
\author{Heinrich Ody
	\institute{Department of Computing Science\\
		University of Oldenburg\\
		Oldenburg, Germany}
	\email{heinrich.ody@uni-oldenburg.de}
}
\def\titlerunning{Monitoring of Traffic Manoeuvres with Imprecise Information}
\def\authorrunning{H. Ody}
 
\def\iscameraready{1} 
\begin{document}

\maketitle 

\input{abstract}

\ifdefined\iscameraready%
\todo[inline]{07.06.: In cameraready I sohuld fix the spacing with units}

\todo[inline]{03.08.: Everywhere $a\in\R$ used for acc, $q\in \Q$ for length and $z\in\R$ for temporal relative delay?}
\fi

\ifdefined\isthesis
\todo{If Iwant to change font of $\spd$ etc. to roman, I have to use different macros for $\spd(C)$ and $\spd_{C,i}$!!!}
\fi

\input{tex/monitoring}


\bibliographystyle{eptcs}
\bibliography{fvav17,manual}

%
%
%
%
%
%

\end{document}

%% file: abstract.tex
\begin{abstract}
In monitoring, we algorithmically check if a single behavior satisfies a property.
Here, we consider monitoring for Multi-Lane Spatial Logic (MLSL).
The behavior is given as a finite transition sequence of MLSL and the property is that a spatial MLSL formula should hold at every point in time within the sequence.
In our procedure we transform the transition sequence and the formula to the first-order theory of real-closed fields, which is decidable, such that the resulting formula is valid iff the MLSL formula holds throughout the transition sequence.
We then assume that temporal data may have an error of up to $\epsilon$, and that spatial data may have an error of up to $\delta$.
We extend our procedure to check if the MLSL formula $\epsilon$-$\delta$-robustly holds throughout the transition sequence.
    \medskip

\end{abstract}

%% file: tex/monitoring.tex
\def\imprecisefigpath{tex/fig/imprecise}
\tikzsetexternalprefix{\imprecisefigpath/}

\section{Introduction}

\emph{Multi-Lane Spatial Logic} (MLSL) comprises an abstract model of a motorway and a spatial logic to reason about traffic configurations \cite{HLOR11}.
MLSL can be used to, e.g.\ analyse controllers of (semi) automated driving systems.

In offline monitoring we are given a recorded behavior $\pi$, a specification $\psi$ and we want to check if $\pi$ satisfies $\psi$, denoted as $\pi \models \psi$.
In this work we perform monitoring for MLSL.
While MLSL has been extended with CTL-like branching time temporal modalities \cite{LH15}, they are not suitable for monitoring.
We formalise what it means for an MLSL formula $\phi$ to hold \emph{globally} in linear time, where we denote `globally $\phi$' as $\ltlglobally \phi$.
This means that here $\pi$ is a transition sequence and we instantiate $\psi$ with $\ltlglobally\phi$, where $\phi$ is an arbitrary MLSL formula.


We define a procedure to check if an MLSL formula holds globally in an MLSL transition sequence.
For this we adapt a procedure to check satisfiability of a restricted form of MLSL formulas \cite{FHO15}.
In this extension we transform the MLSL formula that should hold globally, and the transition sequence to the first-order theory of real closed fields, which is decidable \cite{Tar51} (there called elementary algebra), such that the transformed formula is valid iff $\pi\models \ltlglobally \phi$ holds.

However, it is idealistic to assume that the data we are working with is exact.
Here, we consider errors in positional data (spatial imprecision) and imprecisions of when reservations and claims are set and withdrawn (temporal imprecision).
For temporal robustness other approaches use that they have the satisfaction of temporal atoms as a signal over time.
Here, our temporal formula is $\ltlglobally \phi$ and $\phi$ is our temporal atom.
We do not have the truth value of $\phi$ as a signal over time. 
For this reason we decided to base the temporal aspect of MLSL on \emph{timed words} \cite{AD94}, from which we then derive MLSL transition sequences.
Then we define temporal robustness by deviating the time stamps in a timed word.
We combine this temporal robustness with our previous work on spatial robustness \cite{Ody15} and define spatio-temporal similarity with a metric.
We then define what it means that an MLSL formula globally holds, even if the transition sequence is subject to spatio-temporal perturbations.
Lastly, we extend our previous transformation to accomodate the spatio-temporal perturbations.

%
%

\paragraph{Related Work}
There is a lot of work on monitoring temporal properties in dense time formalisms.
This was then extended to checking how robustly (in the spatial sense) a signal satisfies a Metric Temporal Logic formula \cite{FP09,FP06}.
This was then extended to consider spatio-temporal robustness of Signal Temporal Logic \cite{DM10}, a temporal logic that works with dense time and dense data.
In \cite{FH05} the authors considered robust satisfaction of Duration Calculus.
In all of these works the authors define a multi-valued semantics for their temporal logic.
For MLSL we have not been able to define a useful multi-valued semantics, because the atoms do not have quantitative data, which is crucial in the works mentioned above.
In \cite{AD14} the authors perform online monitoring of spatial properties for a driving car.
In contrast to our work, they take a very low level view (little abstraction) and they can not easily check arbitrary spatial properties.
In \cite{RA15} the authors formalise traffic and traffic rules in a theorem prover.
However, their goal is analysing meta properties, such as ambiguity of traffic rules, rather than automation.
Urban MLSL is an extension of MLSL that allows for logical reasoning about traffic scenarios in an urban setting \cite{HS16,Sch17}.
%
%

\section{Abstract Model for Motorways}

\begin{figure}\centering
	\scalebox{1}{\includegraphics{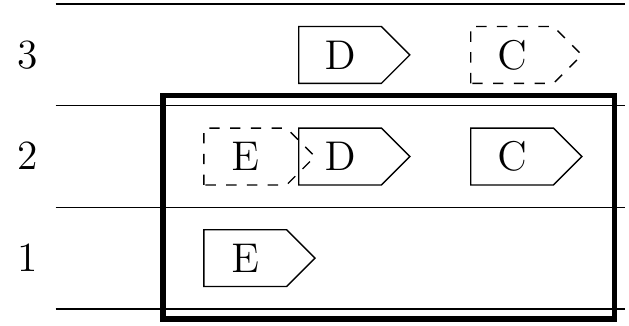}}
	\caption{Visualisation of a traffic snapshot, where car $\carC$ has a reservation (solid line) and a claim (dashed line), car $\carD$ has two reservations and car $\carE$ also has a reservation and a claim.
		The claim of car $\carE$ and a reservation of car $\carD$ overlap.
		Additionally, we show a view (rectangle with thick line).
		Note that one reservation of car $\carD$ and the claim of car $\carC$ are outside of the view.}
	\label{fig:ts1}
\end{figure}

We use an abstract formal model for motorway traffic \cite{HLOR11}, where the traffic configuration at a specific point in time is given by a \emph{traffic snapshot}.
In a traffic snapshot the motorway is represented by two dimensions, a discrete vertical dimension, which represents lanes and a continuous horizontal dimension, which represents the position along a lane.
Then a \emph{reservation} of a car represents space the car physically occupies plus some safety margin, which we assume to be the braking distance.
When a car changes lanes it may have multiple adjacent reservations.
A \emph{claim} of a car represents that the car would like to reserve the claimed space.
With claims we model the turn-signal of a real car.
Additionally, a traffic snapshot has information about the speed and acceleration of each car.
The evolution of traffic over time is modelled as a labelled transition system, where each state is a traffic snapshot.
We give an example traffic snapshot and MLSL formulas to develop some intuition for the formalism.
\begin{example}\label{ex:ts}
	MLSL Formulas are evaluated on a restricted area of a traffic snapshot called \emph{view}.
	We show an example traffic snapshot and view in Figure~\ref{fig:ts1}.
	In the traffic snapshot, with the given view, the formula
	\[
		\somewhere{\free \hchop \re(e) \hchop \free }
	\]
	holds.
	Here, $\somewhere{\cdot}$ is an abbreviation and means that the subformula holds somewhere in the view, $\hchop$ is used to separate adjacent segments within the lane, $\free$ indicates that the lane segment is free of claims and reservations and $\re(e)$ means that the segment has a reservation from car $e$.
	Note that in formulas we use lower case letters to refer to cars.
	The formula 
	\[
		\somewhere{\free \hchop \cl(e) \hchop \re(d) \hchop \free \hchop \re(c) \hchop \free}
	\]
	also is satisfied by the traffic snapshot and the view in Figure~\ref{fig:ts1}.
	With $\cl(e)$ we indicate that the lane segment has a claim of car $e$.
	Note that $\cl(e)$ and $\re(d)$ are not exclusive, i.e.\ in the lane segment where the claim of $\carE$ and the reservation of $\carD$ overlap, both, $\cl(e)$ and $\re(d)$ are satisfied.
	We can stack formulas to express that on the lower lane the lower formula holds, and that on the upper lane the upper formula holds. 
	That is, the formula
	\[
		\vchop*{\free \hchop \re(e) \hchop \free }{\free \hchop \cl(e) \hchop \re(d) \hchop \free \hchop \re(c) \hchop \free}
	\]
	is satisfied with the complete view, not just somewhere within the view.
\end{example}

Let $\Caridentifiers$ be a set of cars and $\Lanes$ be a set of lanes let $\powerset{\Lanes}$ be the powerset over $\Lanes$.
The composition of data from the cars in $\Caridentifiers$ is a \emph{traffic snapshot}.
We add a function $\sensorfunction$, which gives the braking distance of a car, to the traffic snapshot from \cite{HLOR11}.
\begin{definition}[Traffic Snapshot]
	For every car $C$ let $\physicalLength(C)$ be the physical length of $C$.
	Then a traffic snapshot is defined as $\TS = (\pos, \sensorfunction, \spd, \acc, \res, \clm)$, where 
	$\pos: \Caridentifiers \rightarrow \R$ is the position of the rear of a car, 
	$\sensorfunction: \Caridentifiers \rightarrow \R_{> 0}$ is the length of a reservation of a car including its physical length,
	$\spd :\Caridentifiers \rightarrow \R$ is the current speed,
	$\acc : \Caridentifiers \rightarrow \R$ is the current acceleration,
	$\res: \Caridentifiers \rightarrow \powerset{\Lanes}$ is the set of reserved lanes.
	$\clm:\Caridentifiers \rightarrow \powerset{\Lanes}$ is the set of claimed lanes.
\end{definition}
\todo{If Iwant to change font of $\spd$ etc. to roman, I have to use different macros for $\spd(C)$ and $\spd_{C,i}$!!!}

We model the evolution of traffic snapshots as labelled transitions, where we use discrete and continuous transitions.
The discrete transitions for a car $C$ are to change the acceleration ($\setAcceleration(C,a)$ with $a \in \R$), set a claim for a lane ($\setClaim(C,n)$ with $n\in\Lanes$), change an existing claim into a reservation $\setReservation(C)$), withdraw an existing claim ($\withdrawClaim(C)$) and withdraw a reservation from a lane ($\withdrawReservation(C,n)$ with $n\in\Lanes$).
The continuous transitions are similar to delay transitions in timed automata, i.e.\ we update the data affected by time (here position, speed and the derived braking distance).
To define the transitions we use substitution and function overriding, i.e. let $\TS[f/f\subst{C}{x}]$ be $\TS$, except that the function $f$ is replaced by $f\subst{C}{x}$, which maps $C$ to the value $x$ and agrees on everything else with $f$.
\begin{definition}[Transitions]\label{def:transition-sequence}
	Let $n, n' \in \Lanes$ with $n' \in \{n-1,n+1\}$ and $a,z\in\R$.
	Further, to compute the braking distance of a car we assume a maximum deceleration value $\maxdeceleration$ that all cars are capable off.
	We define
	{\allowdisplaybreaks
	\begin{align*}
	\TS \transition{\setAcceleration(C,a)} \TS' &\iff \TS' = \TS[\acc/\acc\subst{C}{a}] \\
	%
	\TS \transition{\setClaim(C,n)} \TS' &\iff \TS' = \TS[\clm/\clm\subst{C}{n}] \land  \res(C) = \{ n'\} \land \clm(C) = \emptyset\\
	%
	\TS \transition{\setReservation(C)} \TS' &\iff \TS' = \TS[\res,\clm/\res\subst{C}{\res(C) \cup clm(C)},\clm\subst{C}{\emptyset}]  \\
	%
	\TS \transition{\withdrawClaim(C)} \TS' &\iff \TS' = \TS[\clm/\clm\subst{C}{\emptyset}]  \\
	%
	\TS \transition{\withdrawReservation(C,n)} \TS' &\iff \TS' = \TS[\res/\res\subst{C}{\{ n \}}]  \land n \in \res(C)\\
	%
	\TS \transition{z} \TS' &\iff \TS' = \TS[\pos,\spd,\sensorfunction / \pos',\spd',\sensorfunction']  \;\mathrm{where}\; \\
		&\qquad \pos' = \{ C \mapsto \pos(C) + \spd(C) \cdot z + \frac{1}{2}\acc(C) \cdot z^2  \mid C \in \Caridentifiers\} \\
		&\qquad \spd' = \{ C \mapsto \acc(C) \cdot z + \spd(C)  \mid C \in \Caridentifiers \} \\
		&\qquad \sensorfunction' = \{ C \mapsto \frac{(\spd(C) + \acc(C) \cdot z)^2}{\maxdeceleration} + \physicalLength(C) | C \in \Caridentifiers\} \qedhere
	\end{align*}
	}
\end{definition}
While we give a definition of $\sensorfunction$ for all cars, our results also hold with a different definition of $\sensorfunction$ for each car.
Such an individual definition could depend on properties of the cars, e.g.\ one definition for light cars and another for heavy cars.
However, our results only hold when the function used is a polynomial, i.e.\ we do not allow exponentiation and trigonometric functions.
Here we take  the view that underlying a transition sequence, there is a \emph{timed word} \cite{AD94}.
A timed word is a sequence of events and time stamps.
\begin{definition}[Timed Words]
	For a set of cars $\Caridentifiers$ and a car $C \in \Caridentifiers$ we denote the cars actions as $\Sigma_C = \{ \setClaim(C,n), \setReservation(C), \removeClaim(C), \removeReservation(C,n), \setAcceleration(C,a) \mid n \in \Lanes, a \in \R \}$ and the set of actions of all cars as $\Sigma = \bigcup_{C \in \Caridentifiers} \Sigma_C$.
	%
	%
	The joint behavior of the cars in $\Caridentifiers$ is a timed word $\timedword = (\sigma, \tau)$ where $\sigma \in \Sigma^*$ and $\tau$ is a weakly monotonic increasing sequence of time stamps over $\R_{\geq 0}$.
	We assume that all timed words have as their last element in $\sigma$ a special marker $\wordend \not\in \Sigma$.
	For a timed word $\rho=(\sigma,\tau)$ with $\sigma=\sigma_1 \dots \sigma_n$ and $\tau = \tau_1 \dots \tau_n$ we denote the \emph{projection} to $\Sigma' \subseteq \Sigma$ as $\rho \projection \Sigma'=(\sigma',\tau')$ with $\sigma'=\sigma_{i_1} \dots \sigma_{i_k}$, $\tau'=\tau_{i_1} \dots \tau_{i_k}$ and $1\leq i_1,\dots, i_k\leq n$ such that $\sigma'$ is the longest subsequence of $\sigma$ that only has letters from $\Sigma' \cup \{ \wordend \}$.
	Let the $\timespan(\timedword)$ of a timed word $\timedword$ be the interval $[0,\tau_n]$.
	We define the \emph{time-bounded prefix} $\timedword_t$ with $t \in \timespan(\timedword)$ as $(\sigma_1, \tau_1)  \dots (\sigma_i, \tau_i) (\wordend, t)$, where $i$ is the largest index such that $\tau_i \leq t$.
	Note that we might have $\tau_i = t$.
\end{definition}

We define that the application of a timed word to a traffic snapshot gives a transition sequence.
The idea is that we first let time advance to the $i$th time stamp and then perform the $i$th discrete action.
Note that we interpret `$\wordend$' as a delay of zero time.
\begin{definition}[From Timed Words to Transition Sequences]
	Given a timed word $\rho=(\sigma,\tau)$ with $\sigma=\sigma_1 \dots \sigma_{n-1}\wordend$ and $\tau = \tau_1 \dots \tau_n$ and a traffic snapshot $\TS_1$, we define the transition sequence $\timedword(\TS_1)$ as
	\[
		\TS_1 \transition{\tau_1} \TS_2
		\transition{\sigma_1} \TS_3
		\transition{\tau_2 - \tau_1} \dots 
		\transition{\tau_{n-1} - \tau_{n-2}} \TS_{2n-2}
		\transition{\sigma_{n-1}} \TS_{2n-1}
		\transition{\tau_{n} - \tau_{n-1}} \TS_{2n}
		\transition{0} \TS_{2n} \enspace.
	\]
	Further, for $t\in\timespan(\timedword)$ we define the \emph{time-bounded transition sequence until $t$} as $\timedword_t(\TS)$ and we denote the last traffic snapshot in $\timedword_t(\TS)$ as $\timedword(\TS)@t$, i.e.\ $\timedword(\TS)@t$ is the traffic snapshot at time $t$.
\end{definition}
In the rest of this work we will only consider transition sequences that result from timed words, and that satisfy the constraints from Definition~\ref{def:transition-sequence}.
Additionally, we assume that all transitions labelled with $\setReservation(C), \removeClaim(C), \removeReservation(C,n)$ change the state, i.e.\ a car makes a reservation only if it has a claim, it withdraws a claim only when it has a claim and it withdraws a reservation only if it has two reservations. 
%

\begin{figure}
	\begin{minipage}{.2\linewidth}
		\centering
		\tikzsetnextfilename{ts-sequence1-new}
		\scalebox{.6}{\includegraphics{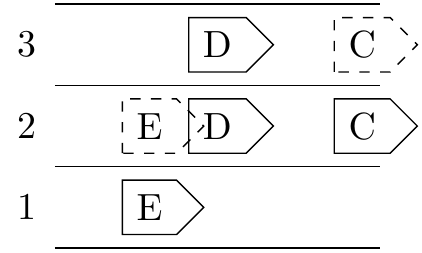}}\\
		{\scriptsize $\TS_1$}
	\end{minipage}
	{\footnotesize$\transition{1\second}$}%
	\begin{minipage}{.2\linewidth}
		\centering
		\tikzsetnextfilename{ts-sequence2-new}
		\scalebox{.6}{\includegraphics{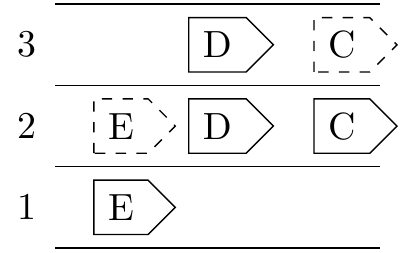}}\\
		{\scriptsize$\TS_2$}
	\end{minipage}
	{\footnotesize$\transition{\removeReservation(\carD,3)}$}%
	\begin{minipage}{.2\linewidth}
		\centering
		\tikzsetnextfilename{ts-sequence3-new}
		\scalebox{.6}{\includegraphics{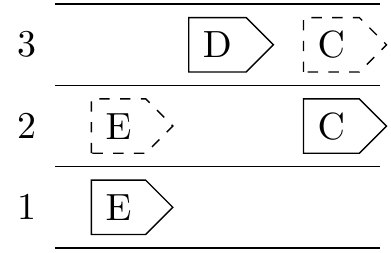}}\\
		{\scriptsize$\TS_3$}
	\end{minipage}
	{\footnotesize$\transition{0.1\second}$}%
	\begin{minipage}{.2\linewidth}
		\centering
		\tikzsetnextfilename{ts-sequence4-new}
		\scalebox{.6}{\includegraphics{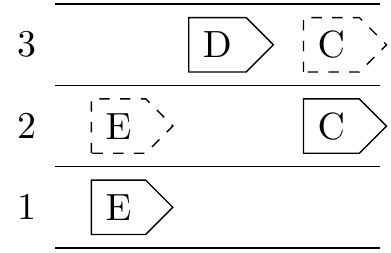}}\\
		{\scriptsize$\TS_4$}
	\end{minipage}\\[1em]%
	{\footnotesize$\transition{\setReservation(\carE)}$}%
	\begin{minipage}{.2\linewidth}
		\centering
		\tikzsetnextfilename{ts-sequence5-new}
		\scalebox{.6}{\includegraphics{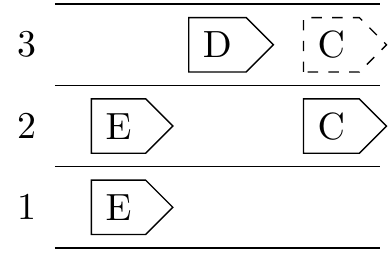}}\\
		{\scriptsize$\TS_5$}
	\end{minipage}%
	{\footnotesize$\transition{5\second}$}%
	\begin{minipage}{.2\linewidth}
		\centering
		\tikzsetnextfilename{ts-sequence6-new}
		\scalebox{.6}{\includegraphics{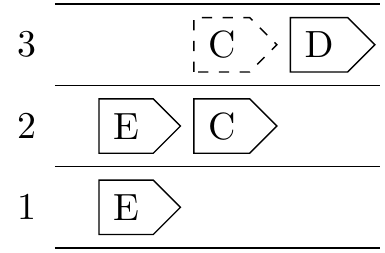}}\\
		{\scriptsize$\TS_6$}
	\end{minipage}%
	{\footnotesize$\transition{\removeReservation(\carE,2)}$}
	\begin{minipage}{.2\linewidth}
		\centering
		\tikzsetnextfilename{ts-sequence6}
		\scalebox{.6}{\includegraphics{\imprecisefigpath/ts-sequence6-new}}\\
		{\scriptsize$\TS_7$}
	\end{minipage}%
	{\footnotesize$\transition{0\second}$}
	\begin{minipage}{.2\linewidth}
		\centering
		\tikzsetnextfilename{ts-sequence6}
		\scalebox{.6}{\includegraphics{\imprecisefigpath/ts-sequence6-new}}\\
		{\scriptsize$\TS_7$}
	\end{minipage}%
	\caption{Visualisation of the transition sequence in Example~\ref{ex:timed-word}.
		Claims are shown with dashed and reservations with solid lines.
		We do not show the last $0\second$ transition to save space}
	\label{fig:imprecise:transition-sequence}
\end{figure}

We give an example of a timed word and how we create a transition sequence from it.
In our examples we give constants representing physical quantities always with their units, i.e.\ $\metre$ for distances, $\second$ for time, $\mps$ for speed, and $\mpss$ for acceleration.
\begin{example}\label{ex:timed-word}
	Let us assume that the global, maximal deceleration constant is given as $\maxdeceleration = 12 \mpss$ and that each car has a physical length of $3\metre$.
	Consider a timed word
	\[
	\timedword = (\removeReservation(\carD,3),1\second)\ (\setReservation(\carE),1.1\second)\ (\removeReservation(\carE,2),6.1\second) \ (\wordend,6.1\second) 
	\]
	and a traffic snapshot $\TS=(\pos,\sensorfunction,\spd,\acc,\res,\clm)$ defined as 
	\begin{flalign*}
	\pos & = \{ \carC \mapsto 60\metre, \carD \mapsto 16\metre, \carE \mapsto 6\metre \} 
			& \acc & = \{ \carC \mapsto 0\mpss, \carD \mapsto 0\mpss, E \mapsto 0\mpss \} \\
	\sensorfunction &= \{ \carC \mapsto 6\metre, \carD \mapsto 30\metre, \carE \mapsto 15\metre \} 
			& \res & = \{ \carC \mapsto \{2\}, \carD \mapsto \{2,3\}, \carE \mapsto \{1\} \} \\
	\spd & = \{ \carC \mapsto 6\mps, \carD \mapsto 18\mps, \carE \mapsto 12\mps \} 
			& \clm & = \{ \carC \mapsto \{3\}, \carD \mapsto \emptyset, \carE \mapsto \{2\} \} 
	\end{flalign*}
	Note that $\TS$ is a formulisation of the traffic snapshot from Figure~\ref{fig:ts1}.
	By applying $\timedword$ to $\TS$, we get the transition sequence 
	\begin{multline*}
	\timedword(\TS) = \TS \transition{1\second} 
	\TS_2 \transition{\removeReservation(\carD,3)} 
	\TS_3 \transition{0.1\second} 
	\TS_4 \transition{\setReservation(\carE)} 
	\TS_5 \transition{5\second} 
	\TS_6 \transition{\removeReservation(\carE,2)}
	\TS_7 \transition{0\second} 
	\TS_7 \transition{0\second} \TS_7
	\end{multline*}
	depicted in Figure~\ref{fig:imprecise:transition-sequence}.
	Note that the two $0\second$ delays in the timed word above result from the delay between $\removeReservation(\carE,2)$ and `$\wordend$', and from our representation of `$\wordend$' in the transition sequence as a $0\second$ delay.
	We use this transition sequence as our running example in this work.
\end{example}

In MLSL we reason about traffic configurations from the local perspective of a car, called \emph{view}.
\begin{definition}[View]
	A view is a tuple $V=(L,X,E)$, where $L=[l,n] \subseteq \Lanes$ is the interval of visible lanes, $X = [r,t] \subseteq \R$ is the extension of the visible lanes and $E \in \Caridentifiers$ is the owner of the view.
	We say that $V' = (L', X', E)$ is a subview of $V$ if $L' \subseteq L$ and $X' \subseteq X$, where we interpret $[l',n'] = \emptyset$ if $l' > n'$ and $\emptyset \subseteq L$ for any set $L$.
	We define $V^{L'} = (L', X, E)$ and $V_{X'} = (L, X', E)$.
\end{definition} 
Additionally, we assume a set of variables $\Carvariables$ ranging over $\Caridentifiers$, a special variable $\ego$ and a variable valuation $\val : \Carvariables \cup \{ \ego \} \rightarrow \Caridentifiers$ such that $\val(\ego) = E$.
We define that $M= (\TS,V,\val)$ is an MLSL model.

We lift transition sequences over traffic snapshots to transition sequences over models, as in \cite{LH15} by moving the view along with its owner.
Let $\timedword$ be a timed word and $\TS_1$ a traffic snapshot with $\timedword(\TS_1) = \TS_1 \transition{\lambda_1}  \dots \transition{\lambda_{n-1}} \TS_n$, where $\lambda_1,\dots, \lambda_{n-1} \in \Sigma\cup \R$.
For $M_1 = (\TS_1, V_1, \val)$, $V_1= (L,[r_1,t_1],E)$ we define 
\[
	\timedword(M_1) = M_1  \transition{\lambda_1}  \dots \transition{\lambda_{n-1}} M_n
\]
with $M_i = (\TS_i, V_i, \val)$, $V_i=(L,[r_i,t_i],E)$ and $r_{i+1} = r_1 + (\pos_{i+1}(E) - \pos_{i}(E)), t_{i+1} = t_1 + (\pos_{i+1}(E) - \pos_{i}(E))$.

The syntax of MLSL is 
\begin{align*}
\phi &::= \gamma=\gamma' \mid \free\mid \re(\gamma) \mid
\cl(\gamma) \mid \length = q \mid \lnot \phi \mid \phi \land \phi \mid \exists c \qsep \phi 
\mid \phi \hchop \phi \mid \vchop{\phi}{\phi}
\end{align*}
where $\gamma,\gamma'\in \Carvariables\cup \{\ego\}$ and $q\in\Q$.
We denote the set of all MLSL formulas with $\Phi$.
We briefly sketch the idea of the logic:
The atom $\re(\gamma)$ (resp. $\cl(\gamma)$) is satisfied when the current view is filled by the reservation (resp. claim) of the car that $\gamma$ points to.
The atom $\free$ is satisfied if the current view does not have a subview where $\re(\gamma)$ or $\cl(\gamma)$ is satisfied for any car and $\length = q$ is satisfied if the extension of the current view has length $q$.
The horizontal chop $\phi_1 \hchop \phi_2$ (resp. vertical chop $\vchop{\phi_1}{\phi_2}$) is satisfied if we can cut the current view into two horizontally (resp. vertically) adjacent subviews on which $\phi_1$ and $\phi_2$ are satisfied.
In the semantics of the vertical chop operator we follow \cite{FHO15}, i.e.\ we distinguish whether the view contains any lanes before chopping.
\begin{definition}[Semantics] \label{def:semantics-MLSLS}
	Let $c\in \Carvariables$, $q\in\Q$ and $\gamma,\gamma'\in\Carvariables\cup\{\ego\}$.  
	Given  a traffic snapshot $\TS$, a view 
	$V = ([l,n],[r,t], E)$ and a valuation $\val$ with $\val(\ego) = E$ 
	we define the \emph{satisfaction}
	of a formula by a model $\mymodel=(\TS,V,\val)$ as follows:
{\allowdisplaybreaks 
	\begin{align*} 
	\mymodel &\models \gamma=\gamma' & \metaequiv & \phantom{=}\val(\gamma) = \val(\gamma') \\
	\mymodel &\models\free &\metaequiv& \phantom{=} (l \not\in \laneReservation(C)\cup \laneClaim(C) \text{ or } [\pos(C), \pos(C) + \sensorfunction(C)] \cap (r,t)=\emptyset)\\
		&&& \phantom{=} \text{for every }C \in \Caridentifiers, \text{ and } l=n \text{ and } r<t \\
	\mymodel &\models \re(\gamma) & \metaequiv & \phantom{=}  
		l \in \laneReservation(\val(\gamma))\text{ and }
		[r,t] \subseteq [\pos(\val(\gamma)), \pos(\val(\gamma)) + \sensorfunction(\val(\gamma))] \text{ and } 
		l=n\text{ and }  r<t\\ 
	\mymodel &\models \cl(\gamma) & \metaequiv & \phantom{=} 
		l \in \laneClaim(\val(\gamma))\text{ and } 
		[r,t] \subseteq [\pos(\val(\gamma)), \pos(\val(\gamma)) + \sensorfunction(\val(\gamma))]
	\text{ and } l=n\text{ and } r<t \\ 
	\mymodel &\models \length=q & \metaequiv & \phantom{=} t-r=q \\
	\mymodel &\models \lnot \phi &\metaequiv & \phantom{=} \mymodel \not\models \phi \\
	\mymodel &\models \phi_1 \land \phi_2 &\metaequiv & \phantom{=} \mymodel\models\phi_1 \text{ and } \mymodel\models\phi_2 \\
	\mymodel &\models \exists c \qsep \phi & \metaequiv&\phantom{=} (\TS, V, \val\subst{c}{C}) \models \phi,   \text{ for some } C \text{ in } \Caridentifiers \\
	\mymodel &\models \phi_1 \hchop \phi_2 & \metaequiv &\phantom{=} 
	(\TS,V_{[r,s]},\val)\models \phi_1 \text{ and } (\TS,V_{[s,t]},\val)\models \phi_2, \\
	& & & \phantom{=} \text{for some } s, \text{ where }r \le s \le t \\[-7pt]
	\mymodel &\models \vchop{\phi_1}{\phi_2} &\metaequiv&\phantom{=} 
	l\leq n \text{ implies }\\[-7pt]
	&&&\phantom{=} \quad (\TS,V^{[l,m]},\val)\models \phi_1 \text{ and } (\TS,V^{[m+1,n]},\val) \models\phi_2\\
	&&&\phantom{=} \quad \text{for some $m$, where $l-1\leq m\leq n$}, \text{ and }\\ 
	&&&\phantom{=}l > n\text{ implies } (\TS,V,\val)\models \phi_1 \text{ and } (\TS,V,\val) \models \phi_2
		\qedhere
	\end{align*}	
}
\end{definition}
We use common abbreviations like $\true$, $\false$, $\lor$ and $\forall$.
For an MLSL formula $\phi$ we also use the spatial \emph{somewhere} modality from \cite{HLOR11} that is defined as
\[
	\somewhere{\phi} \equiv \true \hchop \vchopp{\true}{\phi}{\true} \hchop\true \enspace.
\]

\section{Monitoring Globally Properties}\label{sec:transformation-globally}

In this section we first formalise for an MLSL model $M$, a timed word $\timedword$ and an MLSL formula $\phi$ what the statement `$\phi$ holds globally in $\timedword(M)$' means.
The intuition is that we check for every point in time $t$ within the time span of $\timedword$, whether the model in the transition sequence $\timedword(M)$ at time $t$ satisfies $\phi$, which in symbols is $\timedword(M)@t\models \phi$.
Afterwards, we define a transformation that takes as inputs $\timedword$, $M$ and $\phi$, and creates a formula $\psi \equiv \psi_M \land \psi_\timedword \implies \psi_\phi$ from the first-order theory of real-closed fields \cite{Tar51} (there called elementary algebra). 
In our transformation we mimic the afore mentioned intuition.
The general idea is that $\psi_M$ represents the initial model, $\psi_\timedword$ changes the transformed initial model and $\psi_\phi$ is checked on the changed model.
We use a universally quantified variable $t_\rmf$ and freeze the transformed  model 
at the time given by the value assigned to $t_\rmf$ and discard later changes.
Then we check whether $\psi_\phi$ holds in the frozen model.
As $t_\rmf$ is universally quantified and ranges over the time span of the timed word $\timedword$, $\psi$ is valid iff $\phi$ holds globally in $\timedword(M)$.

In \cite{LH15} the authors extend MLSL with branching CTL-like temporal modalities.
As branching time modalities are not suited for monitoring, we define a linear time \emph{globally} modality, which is satisfied if the subformula is satisfied at every point in time.
\begin{definition}[Global Satisfaction]
	A transition sequence $\timedword(M)$ globally satisfies a spatial property $\phi$ (denoted as $\timedword(M) \models_\mathrm{seq} \ltlglobally \phi$) iff at every point in time $t$ within the span of $\timedword$ the formula $\phi$ is satisfied.
	Formally, 
	\[
	\timedword(M) \models_\mathrm{seq} \ltlglobally \phi \metaequiv \forall t \in \timespan(\timedword) \qsep \timedword(M)@t \models \phi \enspace. \qedhere
	\]
\end{definition}

We consider formulas from first-order theory of real-closed fields with the signature $\left\langle \R, +, \cdot , 0, 1, <,\right\rangle$ and standard interpretation.
The satisfiability problem of this logic is decidable \cite{Tar51}.
We denote the set of all formulas as $\Psi$ and the set of real-valued variables as $\Realvariables$.
This logic shares symbols with MLSL, such as $=$, $\lnot$ and $\land$.
\begin{wrapfigure}[11]{R}{.4\linewidth}
	\centering
	\tikzsetnextfilename{vars}
	\includegraphics{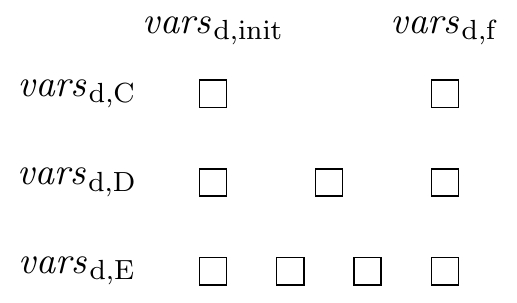}
	\caption{Visualisation of $\vars_\rmd$ structure for the timed word from Example~\ref{ex:timed-word}.
		Only the first and the last column represent the system at the same point in time}
	\label{fig:vars}
\end{wrapfigure}
However, from the context it will be clear to which logic symbols belong.
We denote the \emph{variable assignment} with $\assignment{\cdot}$, which assigns variables a value.

In our transformation the state of a car $C$ at a time point is given by the variables from $\Realvariables$ in the tuple $(\res_{C,i},\res_{C,i}',\pos_{C,i},\sensorfunction_{C,i},\clm_{C,i},\acc_{C,i},\spd_{C,i})$.
For any car $C$ and timed word $\timedword$ let $\timedword_C = (\sigma_C, \tau_C) = \timedword \projection \Sigma_C$.
Now, let $\vars_\rmd$ (d for data) be a list of length $|\Caridentifiers|$ such that it has for each car $C$ a list of length $|\sigma_C| + 1$ and at $\vars_\rmd(C)(i)$ we have an aforementioned tuple of variables.
Note that the lists for the cars may be of different lengths.
We refer to the list that has for each car the first (resp. final) entry with $\vars_{\rmd,\mathrm{init}}$  (resp. $\vars_{\rmd,\rmf}$).
\begin{example}
	Consider the timed word $\timedword$ from Example~\ref{ex:timed-word}. 
	Then let 
	\begin{align*}
	\timedword_\carC &= \timedword \projection \Sigma_\carC = (\wordend,6.1\second) \\
	\timedword_\carD &= \timedword \projection \Sigma_\carD = (\removeReservation(\carD,3),1\second)(\wordend,6.1\second) \\
	\timedword_\carE &= \timedword \projection \Sigma_\carE = (\setReservation(\carE),1.1\second) (\removeReservation(\carE,2),6.1\second) (\wordend,6.1\second) 
	\end{align*}
	For $\Caridentifiers= \{ \carD,\carC,\carE \}$ and $\timedword$ we show the structure of $\vars_\rmd$ in Figure~\ref{fig:vars}.
\end{example}


For real-valued variables, which we consider as not assigned, we introduce a special value $\novalue$ such that $\novalue \not\in \Lanes$.
Given a traffic snapshot $\TS$ we assume w.l.o.g. that for all cars $C\in \Caridentifiers$ we have $\res(C)=\{n,\novalue\}$ if $C$ only reserves lane $n\in \Lanes$ and $\clm(C)=\{\novalue\}$ if $C$ does not have a claim.
Further, let $\vars_\rmd$ be globally available.
\begin{definition}[Transforming Initial Models]
	For a traffic snapshot $\TS$ over a set of cars $\Caridentifiers$, let for a car $C$ $n_C\in \Lanes,n_C' \in \Lanes\cup\{\novalue\}$ be the values in the set $\res(C)$ and $n_C'' \in \Lanes\cup\{\novalue\}$ be the value in the set $\clm(C)$.
	With the variables $\vars_{\rmd,\mathrm{init}}$ we define 
	\begin{multline*}
	\transform_\mathrm{init}(\TS) := \bigwedge_{C\in \Caridentifiers} \pos_{C,1} = \pos(C) \land \res_{C,1} = n_C \land \res_{C,1}' = n_C' \land{}   \\ 
			\spd_{C,1} = \spd(C) \land \acc_{C,1} = \acc(C) \land \clm_{C,1} = n_C'' \land \sensorfunction_{C,1} = \sensorfunction(C) \enspace. \qedhere
	\end{multline*}
\end{definition}

For each car $C\in \Caridentifiers$ the transformation of an action ($\transform_\mathrm{act}$)
is split into a transformation of a delay action ($\transform_\mathrm{delay}$) and into a transformation of a discrete action ($\transform_\mathrm{d-act}$).
We point out that we treat the `$\wordend$' marker as an action that does not change anything.
\begin{definition}[Transforming Actions]
	For some car $C\in \Caridentifiers$ and an index $i\in \N$ let $\sigma_i \in \Sigma_C \cup \{ \wordend\}$.
	For $n \in \Lanes, a \in \R$ and a variable $z \in \Realvariables$ indicating a delay we define 
	{\allowdisplaybreaks
	\begin{align*}
	\transform_\mathrm{act}(\sigma_i, z, i,C) &:= 
		\transform_\mathrm{d-act}(\sigma_i, i,C) \land \transform_\mathrm{delay}(z, i,C) \\
	\transform_\mathrm{d-act}(\sigma_i,i,C) & := 
		\begin{cases}
			\clm_{C,i+1} = n \land \id_{C,i}(\res,\res',\acc) 
					& \metaif \sigma_i = \setClaim(C,n)\\
			\res_{i+1}' = \clm_{C,i} \land \clm_{C,i+1} =\novalue \land \id_{C,i}(\res,\acc)
					& \metaif \sigma_i = \setReservation(C) \\
			\res_{C,i+1} = n \land \res_{C,i+1}' = \novalue \land \id_{C,i}(\clm,\acc)
					& \metaif \sigma_i = \removeReservation(C,n) \\
			\clm_{i+1} = \novalue \land \id_{C,i}(\res,\res',\acc)
					& \metaif \sigma_i = \removeClaim(C)\\
			\acc_{C,i+1} = a \land \id_{C,i}(\res,\res',\clm)
					& \metaif \sigma_i = \setAcceleration(C, a)\\
			\id_{C,i}(\res,\res',\clm,\acc)
					& \metaif \sigma_i = \wordend
		\end{cases} \\
		\transform_\mathrm{delay}(z, i,C) & := 
			\pos_{C,i+1} = \pos_{C,i} + \spd_{C,i} \cdot z + \frac{1}{2}\acc_{C,i} \cdot z^2  \land \spd_{C,i+1} = \acc_{C,i} \cdot z + \spd_{C,i} \\
			& \qquad \land \sensorfunction_{C,i+1} = \frac{(\spd_{C,i} + \acc_{C,i} \cdot z)^2}{\maxdeceleration} + \physicalLength(C) 
	\end{align*}}
	with $\id_{C,i}(\res) := \res_{C,i+1} = \res_{C,i}$ and similar for the other variables.
\end{definition}

Now we can define a transformation for the time-bounded prefix $\timedword_{t_\rmf}$, where $\timedword$ is a timed word and $t_\rmf \in \Realvariables$.
The model at time $t_\rmf$ is stored in the variables $\vars_{\rmd,\rmf}$.
To achieve this we ignore all changes after time $t_\rmf$.
To define our transformation of time-bounded prefixes, we  assume a structure $\vars_\rmt$ that has entries $t_{C,i}\in\Realvariables$, similar to $\vars_\rmd$.
For a car $C$ and a projected timed word $\timedword \projection \Sigma_C = (\sigma_C, \tau_C)$ we identify time stamps with variables that have the constraint $t_{C,i+1} = \tau_{C,i}$ and $t_{C,1} = 0$.
\begin{definition}[Transforming Timed-Bounded Transition Prefixes]
	\label{def:transforming-time-bounded-transition-prefixes}
	For $C \in \Caridentifiers$ let $\timedword = (\sigma,\tau)$, $\timedword \projection \Sigma_C = (\sigma_C, \tau_C)$ and $t_\rmf \in \Realvariables$.
	Then we define 
	\begin{align}		
	\transform_{\mathrm{word}}(\sigma, t_\rmf) & := 
		\bigwedge_{C \in \Caridentifiers} 
		\transform_{C\mathrm{-word}}(\sigma_C, t_\rmf,C) \nonumber\\
	%
	\transform_{C\mathrm{-word}}(\sigma, t_\rmf, C) 
		& := \bigwedge_{i \in \{ 1, \dots, |\sigma| \} } \nonumber \\
		&\qquad  (t_{C,i} \leq t_{C,i+1} \leq t_\rmf \implies \transform_\mathrm{act}(t_{C,i+1} - t_{C,i},\sigma_i, i,C)) \label{eq:transform-both}\\ 
		&\qquad \land  (t_{C,i} \leq t_\rmf < t_{C,i+1} \implies \transform_\mathrm{act}(t_\rmf - t_{C,i},\wordend, i,C)) \label{eq:until-freeze} \\
		&\qquad  \land  (t_\rmf < t_{C,i} \implies \transform_\mathrm{act}(0\second,\wordend, i,C)) \label{eq:freeze} 
	%
	\end{align}
\end{definition}
The first implication \eqref{eq:transform-both} considers the case where the effect of the action takes place before or at time $t_\rmf$.
Hence, we completely represent the effect in our transformation.
If the condition of the second implication \eqref{eq:until-freeze} is satisfied, we know that delaying by $t_{C,i+1} - t_{C,i}$ time units takes us past $t_\rmf$.
Hence, we only delay by $t_\rmf-t_{C,i}$ time units, exactly to time point $t_\rmf$ and do not transform $\sigma_i$.
Instead of $\sigma_i$ we transform `$\wordend$', which ensures that all variables retain their values.
The third implication \eqref{eq:freeze} ensures that we do not manipulate the model anymore, after time point $t_\rmf$.
Note that in the conditions of the implications for each $i$ exactly one condition is satisfied.


We need a method to check if an MLSL model satisfies an MLSL formula.
In \cite{FHO15} the authors defined a transformation to check satisfiability of an MLSL formula $\phi$ that is restricted to a finitely bounded set of cars (called well-scoped MLSL with scopes).
Their transformation creates a quantified linear integer-real arithmetic formula that is valid iff $\phi$ is satisfiable.
We simplify their transformation to instead check whether for a \emph{given model} $M$ it holds that $M\models \phi$.
The adapted transformation takes two parameters: the first is a tuple $\Upsilon = (\mathit{CS}, l, n, x_{\rmf,\rml}, x_{\rmf,\rmr}, \val)$, defining the cars to consider (here we have $\mathit{CS} = \Caridentifiers$), the current lanes $[l,n]$ with $l,n \in \N$, the current extension as variables $x_{\rml}, x_\rmr \in \Realvariables$, and the valuation function $\val$.
The second parameter is the MLSL formula.
The formula that $\transform_\rmf$ creates is from the first-order theory of real-closed fields and represents the semantics of MLSL.
For this the formula creates suitable constraints on $\vars_{\rmd,\rmf}$.
Note that negation in MLSL is represented with $\transform_\rmf$ by negation in the first-order theory of real-closed fields, i.e.\ for all $\Upsilon$ and MLSL formulas $\phi$ we have $\transform_\rmf(\Upsilon,\lnot \phi) = \lnot\transform_\rmf(\Upsilon,\phi)$.
The following claim states that we can algorithmically determine if an MLSL formula is satisfied by a model.
\begin{claim}\label{lem:formula-transformation}
	Let $M = (\TS,V,\val)$ with $V=([l,n],[r,t],E)$ and let for $\Caridentifiers$ the variables $\vars_{\rmd,\mathrm{init}}$ be available. 
	We constrain $x_\rml,x_\rmr \in \Realvariables$ with $x_\rml = r, x_\rmr = t$ and define  $\Upsilon=(\Caridentifiers,l,n,x_\rml,x_\rmr,\val)$. 
	Then for any MLSL formula $\phi$ we have
	\[
		\transform_\mathrm{init}(\TS) \implies \transform_\rmf(\Upsilon,\phi) \text{\rm{} is valid} \qquad \metaequiv\qquad  M \models \phi \enspace,
	\]
	where $\transform_\rmf(\Upsilon,\phi)$ is evaluated on the variables $\vars_{\rmd,\mathrm{init}}$.
\end{claim}


Now we can define our transformation to check globally properties.
The intuition of the transformation is that it checks if we can stop the evolution of $\timedword(M)$ at all time points  $t_\rmf$ and store the model at that time in the variables subscripted with `f' and then evaluate $\phi$ on this stored model. 
Note that we use the variables $x_{\rmf,\rml},x_{\rmf,\rmr}$ to represent the extension at time $t_\rmf$.
\begin{definition}[Transforming Globally Properties]
	Given a model $M$ and a timed word $\timedword$ over a finite set $\Caridentifiers$ we use the variables $x_{\rmf,\rml}, x_{\rmf,\rmr} \in \Realvariables$ with the constraints $x_{\rmf,\rml} = r + (\pos_{E,\rmf} - \pos_{E,1})$ and $x_{\rmf,\rmr} = t + (\pos_{E,\rmf} - \pos_{E,1})$.
	Let $\Upsilon = (\Caridentifiers, l, n, x_{\rmf,\rml}, x_{\rmf,\rmr}, \val)$, then for an MLSL formula $\phi$ we define
	\begin{align*}
	\transform_\ltlglobally(\timedword, M, \phi) &:= 
			\forall t_\rmf \in \timespan(\timedword) \qsep  (\transform_\mathrm{word}(\sigma,t_\rmf) \land \transform_\mathrm{init}(\TS))
			\implies \transform_\mathrm{f}(\Upsilon, \phi) \enspace, 
	\end{align*}
	where $\transform_\mathrm{f}(\Upsilon, \phi)$ is evaluated over $\vars_{\rmd,\rmf}$.
\end{definition}

\begin{claim}\label{lem:globally-phi-equivalent-transformation}
Given a timed word $\timedword$, an MLSL model $M$ and an MLSL formula $\phi$ 
\[
	\timedword(M) \models_\mathrm{seq} \ltlglobally \phi \metaequiv \transform_\ltlglobally(\timedword, M, \phi) \text{ is valid}\enspace.
\]
\end{claim}
The previous claim states that we can reduce checking $\timedword(M) \models_\mathrm{seq} \ltlglobally \phi$ to checking $\transform_\ltlglobally(\timedword, M, \phi)$ for validity.
This is equivalent to $\lnot\transform_\ltlglobally(\timedword, M, \phi)$ being unsatisfiable.
As the satisfiability of first-order theory of real closed fields is decidable \cite{Tar51}, we get the following theorem, assuming that the above claim holds.
\begin{theorem}
	It is decidable whether an MLSL formula holds globally in an MLSL transition sequence.	
\end{theorem}


\begin{example}\label{ex:transformation-potential-collision}
Consider the timed word $\timedword$ and the traffic snapshot $\TS$ from Example~\ref{ex:timed-word} and the MLSL formula \emph{no potential collision} 
\[
	\mathrm{npc} \equiv \forall c,c' \qsep c \neq c' \implies \lnot\somewhere{(\cl(c) \lor \re(c)) \land (\cl(c') \lor \re(c')) }\enspace,
\]
which is a generalisation of the potential collision formula from \cite{HLOR11}.
The formula $\mathrm{npc}$ states that nowhere in the current view, there is an overlap of the claims or reservations from two different cars.
Let the view be $V=([1,3],[0,90],\carE)$ and the valuation be $\val = \{\ego \mapsto \carE\}$, then we define $M=(\TS, V, \val)$.
We give an overview of how our procedure works to find that $\timedword(M) \models_\mathrm{seq} \ltlglobally \mathrm{npc}$ does not hold.

To test whether `there is never a potential collision' holds  in $\timedword(M)$ we check $\transform_\ltlglobally(\timedword, M, \mathrm{npc})$ for validity.
We show that $\transform_\ltlglobally(\timedword, M, \mathrm{npc})$ is not valid by giving a satisfying assignment for its negation.
The negation $\lnot\transform_\ltlglobally(\timedword, M, \mathrm{npc})$ evaluates to 
\begin{align*}
\exists t_\rmf \in \timespan(\timedword) \qsep  
	\transform_\mathrm{word}(\sigma, t_\rmf)
	\land \transform_\mathrm{init}(\TS)) 
	\land \lnot \transform_\mathrm{f}(\Upsilon, \mathrm{npc}) \enspace.
\end{align*}
\todo{perhaps not use $\lnot\transform_\rmf(\Upsilon,\mathrm{npc}) = \transform_\rmf(\Upsilon,\lnot\mathrm{npc})$ at all?Does it help understanding?}

The formula $\mathrm{npc}$ is violated already in the initial model, because the claim of $\carE$ overlaps with the reservation of $\carD$ (cf.\ Figure~\ref{fig:imprecise:transition-sequence}).
However, to give a better insight into our construction we choose to show that $\mathrm{npc}$ is violated \emph{during} the transition from $\TS_5$ to $\TS_6$, at time $\assignment{t_\rmf} = 4\second$.

We show the constraints generated by $\transform_\mathrm{word}(\sigma,4\second)$ for the cars $\carC,\carD$ in Figure~\ref{fig:word-transformations}.
We see that the position of $\carC$ at time $\assignment{t_\rmf} = 4\second$ is its initial position, plus the distance covered in $4\second$, i.e.\ $\assignment{\pos_{C,2}} = 84\metre$.
For car $\carD$ we see that at time $\assignment{t_{\carD,2}} = 1\second$ the withdrawal of a reservation is performed and that the position of $\carD$ is updated to $\assignment{\pos_{\carD,2}} = 34\metre$.
Then, at time $\assignment{t_\rmf} = 4\second$ car $\carD$ is moved for $3\second$ multiplied with its speed to $\assignment{\pos_{\carD,3}} = 88\metre$.

We have $\lnot\transform_\rmf(\Upsilon,\mathrm{npc}) = \transform_\rmf(\Upsilon,\lnot\mathrm{npc})$.
The formula $\lnot\mathrm{npc}$ is evaluated on $\vars_{\rmd,\rmf}$ and the view updated by the movement of $\carE$.
After $4\second$ car $\carE$ has moved $12\mps \cdot 4\second = 48\metre$.
Hence, the updated left and right extension of the view are $\assignment{x_{\rml,\rmf}} = 48\metre$ and $\assignment{x_{\rmr,\rmf}} = 138\metre$.
Now we can check if $\lnot\mathrm{npc}$, which states that there is a subview where the claims or reservations of two different cars overlap, is satisfied.
As $\carC$ has a braking distance of $\assignment{\sensorfunction_{\carD,3}} = 6 \metre$ it claims the interval $[84\metre, 84\metre+6\metre]$ on lane $3$.
As car $\carD$ has a reservation on lane $3$ and its position is within $[84\metre,90\metre]$ the claim of $\carC$ and the reservation of $\carD$ overlap. 
Thus, we have shown $\timedword(M) \not\models_\mathrm{seq} \ltlglobally \mathrm{npc}$. \qedhere
\begin{figure}\centering
	\begin{minipage}{.35\linewidth}
		\tikzsetnextfilename{transformation-C}
		\includegraphics{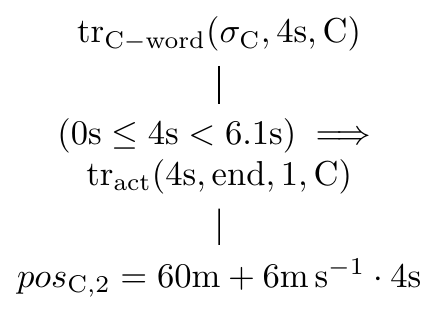}
	\end{minipage}%
	\begin{minipage}{.65\linewidth}
		\tikzsetnextfilename{transformation-D}
		\includegraphics{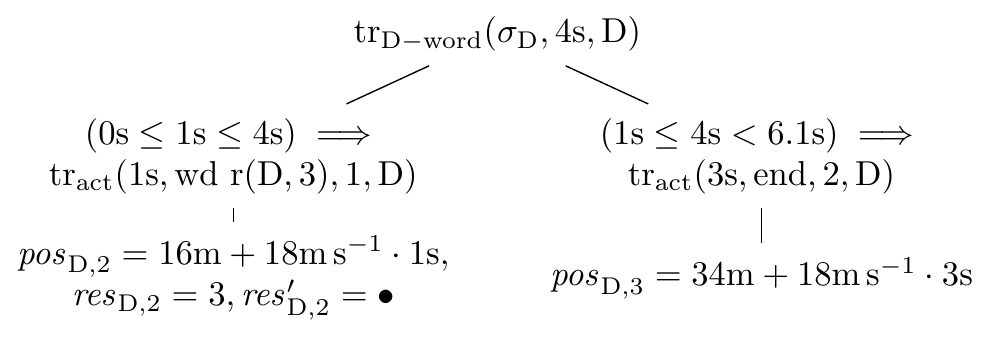}
	\end{minipage}
	\caption{Transformations of the words projected to actions from car $\carC$ and car $\carD$}
	\label{fig:word-transformations}
\end{figure}

\end{example}
\ifdefined\isthesis%
\todo[inline]{What about correctness?? In the file below are some first definitions and proof attempts}
\fi%

\section{Monitoring Globally Properties with Imprecise Information}

In this section we extend our transformation to check, whether an MLSL formula holds globally in a transition sequence with $\epsilon$-$\delta$-robustness.
This allow us to check if, e.g.\ a behavior given as a transition sequence is \emph{barely safe} or if it is \emph{robustly safe}.

Here, we consider errors in positional data and imprecisions of when reservations and claims are set and withdrawn.
Similarity on timed words has originally been defined in \cite{GHJ97}.
However, usually the requirement is imposed that the order of events is equal in similar words.
For distributed systems this requirement seems too strong.
Here, we weaken this requirement and allow the order of independent actions to change in similar words.
For a single car $C$, setting and withdrawing claims and reservations are independent of changing acceleration.
Between different cars, all actions are independent.
We first define an \emph{independence relation} for actions.
\begin{definition}[Independence Relation]
Let $\Sigma_C$ be the action alphabet for car $C$.
We define the \emph{independence relation} as
\begin{align*}
	I_C &= 
		(\{ \setClaim(C, n), \setReservation(C), \removeClaim(C), \removeReservation(C,n) \mid n \in \Lanes \} \times \{ \setAcceleration(C,a) \mid a \in \R \}) \\
		&\qquad\qquad\cup (\{ \setAcceleration(C,a) \mid a \in \R \} \times \{ \setClaim(C, n), \setReservation(C), \removeClaim(C), \removeReservation(C,n) \mid n \in \Lanes \}) \\
	I &= \bigcup_{C, C' \in \Caridentifiers, C \neq C'} (\Sigma_C \times \Sigma_{C'}) \cup \bigcup_{C \in \Caridentifiers} I_C \qedhere
\end{align*}
\end{definition}
We define that two timed words have \emph{equal causality} iff their untimed words are in the same \emph{equivalence class} in the sense of Mazurkiewicz traces \cite{Maz86}.
Two words are in the same equivalence class iff we can create one word from the other by repeatedly swapping letters that are adjacent and independent.
%
%
%
\begin{definition}[Causality Equivalence]
Let $I$ be our independence relation.
For two words $\sigma,\sigma'\in \Sigma^*$ with $\sigma = \sigma_1 \dots \sigma_n$ we define that $\sigma$ and $\sigma'$ are in the same \emph{equivalence class} (denoted $\sigma \in [\sigma']$) as
\begin{align*}
\sigma \in [\sigma'] \metaequiv 
		&\sigma = \sigma' \metaor \text{ there is } \sigma'' \in \Sigma^* \metaand i\in \{1,\dots,n \}  \text{ such that } \\
				&\qquad (\sigma_i,\sigma_{i+1}) \in I \metaand 
				\sigma'' = \sigma_1 \dots \sigma_{i-1} \sigma_{i+1} \sigma_{i} \sigma_{i+2} \dots \sigma_n \metaand
				\sigma''\in [\sigma'] 
\end{align*}
Two timed words $\timedword = (\sigma,\tau),\timedword'=(\sigma',\tau')$ are \emph{causally equivalent} iff $\sigma \in [\sigma']$.
\end{definition}
\begin{example}
Consider $\sigma$ from Example~\ref{ex:timed-word} and $\sigma_1,\sigma_2,\sigma_3$ shown below:
\begin{alignat*}{3}
	 \sigma &= \removeReservation(\carD,3)\ \setReservation(\carE)\ \removeReservation(\carE,2) \ \wordend 
			& \qquad\sigma_1 &= \setReservation(\carE)\ \removeReservation(\carD,3)\ \removeReservation(\carE,2) \ \wordend \\
		\sigma_2 &= \setReservation(\carE)\ \removeReservation(\carE,2) \ \removeReservation(\carD,3)\ \wordend 	
			& \qquad\sigma_3 &= \removeReservation(\carD,3)\ \removeReservation(\carE,2) \ \setReservation(\carE)\ \wordend 	
\end{alignat*}
As $(\removeReservation(\carD,3), \setReservation(\carE)) \in I$ we have $\sigma_1 \in [\sigma]$.
Further, $(\removeReservation(\carD,3)\ \removeReservation(\carE,2)) \in I$ means that $\sigma_2 \in [\sigma_1]$.
By transitivity, this implies $\sigma_2 \in [\sigma]$.
However, $(\removeReservation(\carE,2), \setReservation(\carE)) \not\in I$, which means that $\sigma_3 \not\in [\sigma]$. 
\end{example}

To formalise similarity usually metrics on a set are introduced to define distances between elements.
Then, similarity can be quantified with these metrics.
To capture positional similarity of two models we assign a distance of $\infty$ if any other data than position, extension or sensor function differ.
Otherwise, we assign the maximal difference of these values.
The definition is taken from \cite{Ody15}.
\begin{definition}[Metric on MLSL Models]
Given two models MLSL $M,M'$ we define $\metric_\mathrm{model}(\mymodel,\mymodel') := \infty$ if $\res \neq \res'$ or $\clm \neq \clm'$ or $L \neq L'$ or $\val \neq \val'$ and
\begin{align*}
\metric_\mathrm{model}(\mymodel,\mymodel') &:= 
	\max_{C \in \Caridentifiers  } \{
		|\pos(C) - \pos'(C)|, 
		|\pos(C) + \sensorfunction(C) - (\pos'(C) + \sensorfunction'(C)) |, 
		|r - r'|,
		|t - t'|
	\}
\end{align*}
otherwise.
For $\delta \in \R_{>0}$ we say that two models $M,M'$ are $\delta$-similar, if $\metric_\mathrm{model}(\mymodel,\mymodel') \leq \delta$.
\end{definition}
Additionally, we define a metric on timed words.
We assume that between two similar timed words the time stamps of all acceleration actions are equal.
The reason for this restriction is that if we allow acceleration time stamps to differ, perturbations may accumulate. 
In this work we do not consider such issues.
Furthermore, we require that the time span of two similar timed words is equal.
This is a technical restriction that likely can be removed.
%
\begin{definition}[Metric on Timed Words]
	Let $\Sigma_\setAcceleration = \{\setAcceleration(C,a) \mid C \in \Caridentifiers, a \in \R \}$.  
	Given two timed words $\timedword,\timedword'$ we define $\metric_\mathrm{time}(\timedword,\timedword') = \infty$ if they are not causally equal, $\timedword \projection\Sigma_\setAcceleration \neq \timedword' \projection\Sigma_\setAcceleration$, or if $\timespan(\timedword) \neq \timespan(\timedword')$.
	Otherwise, for a car $C$ let $\timedword_C= (\sigma_C,\tau_C) = \timedword \projection \Sigma_C \setminus \Sigma_\setAcceleration $ and $ \timedword_C'= (\sigma_C',\tau_C') = \timedword' \projection \Sigma_C \setminus \Sigma_\setAcceleration $, where $\sigma_C,  \sigma_C'$ both have length $n_C$.
	We define
	\begin{equation*}
		\metric_\mathrm{time}(\timedword,\timedword') := 
		\max_{C \in \Caridentifiers}\{ 	
			\metric_C(
				\timedword_C,
				\timedword_C'
			)
		\}  \qquad \metaand \qquad
		\metric_C(\timedword_C, \timedword_C') :=
		\max_{i \in \{1, \dots, n_C \} } \{ | \tau_{C,i} - \tau_{C,i}'| \} \enspace. \qedhere
	\end{equation*}
\end{definition}
Note that $\metric_C$ essentially is taken from \cite{GHJ97}.
Further, we point out that because $\timedword,\timedword'$ in the above definition are causally equivalent,  for $\timedword \projection \Sigma_C \setminus \Sigma_\setAcceleration = (\sigma_C, \tau_C)$ and $\timedword' \projection \Sigma_C \setminus \Sigma_\setAcceleration = (\sigma_C', \tau_C')$ we have $\sigma_C = \sigma_C'$.
We lift the metric on timed words to a metric on transition sequences.
For two models $M,M'$ we define $\metric_\mathrm{seq}(\timedword(M),\timedword'(M'))=\infty$ if $M \neq M'$ and otherwise
\[
\metric_\mathrm{seq}(\timedword(M),\timedword'(M')) := \metric_\mathrm{time}(\timedword,\timedword') \enspace.
\]
As for models, two transition sequences are $\epsilon$-similar, if $\metric_\mathrm{seq}$ assigns them a distance $\leq \epsilon$.

\ifdefined\isthesis%
\begin{example}
We give an example of the difficulty arising if we do not use restriction b) from above.
Consider two timed words 
\begin{align*}
&\begin{multlined}[c][.9\displaywidth]
\timedword  = (\setAcceleration(\carE,3\mpss),1), (\setAcceleration(\carC,-3\mpss),1), (\setAcceleration(\carE,-3\mpss),2), (\setAcceleration(\carC,3\mpss),2), \\
(\setAcceleration(\carE,3\mpss),3), (\setAcceleration(\carC,-3\mpss),3), \dots, (\setAcceleration(\carE,-3\mpss),n), (\setAcceleration(\carC,3\mpss),n), (\wordend,n+1)
\end{multlined} \\
&\begin{multlined}[c][.9\displaywidth]
\timedword' = 
(\setAcceleration(\carE,3\mpss),1 + \epsilon), (\setAcceleration(\carC,-3\mpss),1 + \epsilon), (\setAcceleration(\carE,-3\mpss),2 - \epsilon), (\setAcceleration(\carC,3\mpss),2 - \epsilon), \\
(\setAcceleration(\carE,3\mpss),3 + \epsilon), (\setAcceleration(\carC,-3\mpss),3 + \epsilon), \dots, (\setAcceleration(\carE,-3\mpss),n - \epsilon), (\setAcceleration(\carC,3\mpss),n - \epsilon), (\wordend,n+1)
\end{multlined}
\end{align*}
where we interpret time units as seconds and a model $M$ for which we only give speed and acceleration:
\begin{align*}
\spd(\carC) &= 33\mps & \acc(\carC) &= -3\mpss \\
\spd(\carE) &= 30\mps & \acc(\carE) &= 3\mpss
\end{align*}
If we apply $\timedword$ to $M$ at the end they will be the same distance apart as at the start.
More specifically, let $n=7$, then in both words $8\si{\second}$ pass.
Each second both cars travel $31.5\si{\metre}$.
Hence, overall they travel $ 8 \cdot 31.5\si{\metre} =252\si{\metre}$.
However, if we apply $\timedword'$ to $M$ and set $\epsilon = 0.1$.
Then at the end $\carC$ has covered more ground than $\carE$, because the time stretches where $\carE$ is accelerating more than $\carC$ are of length $1-2\epsilon$ and the time stretches where $\carC$ is accelerating more than $E$ are of length $1+2\epsilon$.
In numbers, $\carE$ travelled $268.77\si{\metre}$ and $\carC$ travelled $235.23\si{\metre}$, i.e.\ the difference in travelled distance is about $33\si{\metre}$.
\end{example}
\fi

We define that a model $\delta$-robustly satisfies a formula if all $\delta$-similar models also satisfy the formula.
\begin{definition}[Robust Satisfaction of MLSL Formulas]
Given a model $M$, a desired error allowance $\delta \in \R_{> 0}$ and a formula $\phi$, we define that $M$ satisfies $\phi$ with robustness $\delta$ as 
\[
	M \models^\delta \phi \metaequiv \forall M' \qsep \metric_\mathrm{model}(\mymodel,\mymodel') \leq \delta \implies M' \models \phi \enspace. \qedhere
\]
\end{definition}
We define what it means for a transition sequence to robustly satisfy $\ltlglobally \phi$.
\begin{definition}[Robust Global Satisfaction]
Let $M$ be an initial model, and let $\timedword$ be a timed word.
Then for a formula $\phi$, an allowed spatial error $\delta \in \R_{>0}$ and an allowed temporal error $\epsilon \in \R_{> 0}$, we define
\begin{align*}
\timedword(M) \models_\mathrm{seq}^\delta \ltlglobally \phi &\iff \forall t \qsep t \in \timespan(\timedword) \implies \timedword(M)@t \models^\delta \phi \\
%
\timedword(M) \models_\mathrm{seq}^{\epsilon,\delta} \ltlglobally \phi &\metaequiv \forall \timedword(M)'\qsep \metric_\mathrm{seq}(\timedword(M),\timedword(M)') \leq \epsilon 
\implies \timedword(M)' \models_\mathrm{seq}^\delta \ltlglobally \phi \qedhere
\end{align*} 
\end{definition}

\ifdefined\isthesis%
\todo[inline]{02.06.: For thesis, not for paper. Currently our definitions  have a binary understanding of robustness.
	Perhaps consider $\models^\rmr \phi := \inf\{ d(M,M') | M \models \phi \lnot\iff M' \models \lnot \phi \}$ as a real-valued function, and consider our binary approach as a simplification of the real-valued one?}
\fi

For example, with our notion of similarity the two timed words $\timedword=(\setAcceleration(C,5,1))(\setReservation(C),1.1)$ and $\timedword'=(\setReservation(C),1)(\setAcceleration(C,5,1.1))$ are $0.1$-similar.
However, in our transformation it is cumbersome to consider for a single car different possible sequences of events.
Hence, we assume that two discrete actions of the same car are strictly more than $2\epsilon$ time units apart.
This assumption has the additional benefit that when we consider two $\epsilon$-similar timed words, they are causally equal because all dependent actions have the same order in both timed words.
\begin{assumption}
	For a timed word $\timedword$ and for any car $C$ let $\timedword_C = (\sigma_C, \tau_C) = \timedword \projection \Sigma_C$ and let $\sigma_C$ be of length $n_C$.
	Then, for $\epsilon \in \R_{>0}$ we assume 
	\[
			\min_{C \in \Caridentifiers, i,j \in \{ 1, \dots, n_C \} \;\mathrm{with}\; i \neq j}
					\{ |\tau_{C,i} - \tau_{C,j}| \} > 2\epsilon \enspace. \qedhere
	\]
\end{assumption}
Note that this is not a discretisation, as, e.g.\ actions from different cars may still be arbitrary close.

%
%

Before we can define the transformation we introduce perturbed versions of $\vars_\rmt$ and $\vars_{\rmd,\rmf}$, which we call $\widetilde{\vars}_\rmt$ and $\widetilde{\vars_\rmd}$.
We need to operate on the unperturbed, and the perturbed variables.
To the transformation from Section~\ref{sec:transformation-globally} we add $\tilde{\rmd}$  to indicate that we operate on the perturbed data variables, and $\rmd$ otherwise.
Further, we do not need the unperturbed temporal variables, which we indicate with $\tilde{\rmt}$.
With $\forall \widetilde{\vars_\rmt}$ or $\forall \widetilde{\vars_{\rmd,\rmf}}$ we mean that all variables in $\widetilde{\vars_\rmt}$ or $\widetilde{\vars_{\rmd,\rmf}}$ are universally quantified, and similar for $\exists \widetilde{\vars_\rmt}$ or $\exists \widetilde{\vars_{\rmd,\rmf}}$.
Additionally, we define for $C \in \Caridentifiers, i \in \N$ that $\posSensor_{C,i} = \pos_{C,i} + \sensorfunction_{C,i}$ and $\widetilde{\posSensor}_{C,i} = \widetilde{\pos}_{C,i} + \widetilde{\sensorfunction}_{C,i}$ and for $v,v' \in \Realvariables, r \in \R$ let $v \in v' \pm r = v' - r \leq v \leq v' + r$.

To check if $\phi$ holds $\epsilon$-$\delta$-robustly we first perturb the timestamps of non-acceleration events ($\shake_\epsilon$).
Then we encode the temporally perturbed transition sequence ($\transform_\mathrm{init}^\rmd$ and $\transform_{\mathrm{word}}^{\rmd,\tilde{\rmt}}$), and finally we evaluate the MLSL formula on the perturbed final model ($\shake_{\delta}$ and $ \transform_\rmf^{\tilde{\rmd}}$).
\begin{definition}[Transforming Robust Globally Properties]
For a timed word $\timedword$ and a car $C$ let $(\sigma_C,\tau_C) = \timedword \projection \Sigma_C$ with $\sigma_C$ having length $n_C$, $\Sigma_\setAcceleration = \{\setAcceleration(C,a) \mid a\in\R \}$, $\tilde{x}_{\rml,\rmf}, \tilde{x}_{\rmr,\rmf}\in \Realvariables$ and $\Delta_E = \pos_{E,\rmf} - \pos_{E,1}$.
Then, given a model $M = (\TS,V,\val)$ with $V = ([l,n],[r,t],E)$, a tuple $\Upsilon = (\Caridentifiers, l,n,\tilde{x}_{\rml,\rmf}, \tilde{x}_{\rmr,\rmf}, \val)$ an MLSL formula $\phi$ and $\epsilon,\delta \in \R_{>0}$ we define
\begin{align*}
\transform_\ltlglobally^{\epsilon,\delta}(\timedword, M,  \phi) &:= 
	\forall \widetilde{\vars}_{\rmd,\rmf}, \tilde{x}_{\rml,\rmf}, \tilde{x}_{\rmr,\rmf} \qsep  \forall \widetilde{\vars}_\rmt, t_\rmf \in \timespan(\timedword) \qsep \\
		&\qquad \qquad \transform_\mathrm{init}^\rmd(\TS)
		\land \mathrm{shake}_{\epsilon}(\timedword) 
		\land \mathrm{shake}_{\delta} 
		\land\transform_{\mathrm{word}}^{\rmd,\tilde{\rmt}}(\sigma, t_\rmf) 
		\implies \transform_\rmf^{\tilde{\rmd}}(\Upsilon, \phi)\\
\mathrm{shake}_{\epsilon}(\timedword) & :=
	\bigwedge_{C \in \Caridentifiers, i \in \{ 1,\dots,n_C-1 \} } 
	(\sigma_{C, i} \not\in \Sigma_{\setAcceleration} \implies \tilde{t}_{C,i+1} \in \tau_{C,i} \pm \epsilon) \land (\sigma_{C,i} \not\in \Sigma_{\setAcceleration} \implies \tilde{t}_{C,i+1} = \tau_{C,i})\\
\mathrm{shake}_{\delta} &:=
	\bigwedge_{C \in \Caridentifiers} 
			\widetilde{\pos}_{C,\rmf} \in \pos_{C,\rmf} \pm \delta 
			\land \widetilde{\posSensor}_{C,\rmf} \in \posSensor_{C,\rmf} \pm \delta \land \tilde{x}_{\rml,\rmf} \in (r + \Delta_E) \pm \delta \land {} \\
		&\qquad \tilde{x}_{\rmr,\rmf} \in (t + \Delta_E) \pm \delta \land \widetilde{\clm}_{C,\rmf} = \clm_{C,\rmf}\land \widetilde{\res}_{C,\rmf} = \res_{C,\rmf} \land \widetilde{\res}_{C,\rmf}' = \res_{C,\rmf}' \qedhere
\end{align*}
\end{definition}

\begin{claim}\label{lem:globally-phi-equivalent-robust-transformation}
	Given a timed word $\timedword$ and an MLSL model $M$ and an MLSL formula $\phi$ and $\epsilon,\delta \in \R_{> 0}$ we have 
	\[
		\timedword(M) \models_\mathrm{seq}^{\epsilon,\delta} \ltlglobally \phi \metaequiv \transform_\ltlglobally^{\epsilon,\delta}(\timedword, M,  \phi) \text{ is valid}\enspace.
	\]
\end{claim}
From the previous claim it follows that $\timedword(M) \models_\mathrm{seq}^{\epsilon,\delta} \ltlglobally \phi$ holds iff $\transform_\ltlglobally^{\epsilon,\delta}(\timedword, M,  \phi)$ is valid.
This is equivalent to $\lnot\transform_\ltlglobally^{\epsilon,\delta}(\timedword, M,  \phi)$ being unsatisfiable.
We assume that the above claim holds.
Then, as satisfiability is decidable for the first-order theory of real-closed fields \cite{Tar51}, we get the following theorem.
\begin{theorem}
	For $\epsilon,\delta \in \R_{>0}$ it is decidable whether an MLSL formula holds globally in an MLSL transition sequence with $\epsilon$-$\delta$-robustness.	
\end{theorem}

\begin{example}\label{ex:transformation-robust-safe}
Consider the timed word $\timedword = (\sigma,\tau)$ and traffic snapshot $\TS$ from Example~\ref{ex:timed-word} and the initial model $M=(\TS,V,\val)$ with $V=([1,3],[0,90],\carE)$, $\val = \{\ego \mapsto \carE\}$ from  Example~\ref{ex:transformation-potential-collision}.
We use the formula
\[
	\safe \equiv \forall c, c' \qsep c \neq c' \implies \lnot \somewhere{\re(c) \land \re(c')}
\]
from \cite{HLOR11}, which states that there do not exist  two different cars with overlapping reservations.
In the following let $\epsilon= 0.1\second$ and $\delta = 1\metre$.
To determine that  $\ltlglobally\mathrm{safe}$ does \emph{not} hold $\epsilon$-$\delta$-robustly in $\timedword(M)$, we give a satisfying assignment for the formula $\lnot\transform_\ltlglobally^{0.1\second,1\metre}(\timedword,M,\safe)$.
This formula evaluates to
\[
	\exists \widetilde{\vars}_{\rmd,\rmf}, \tilde{x}_{\rml,\rmf}, \tilde{x}_{\rmr,\rmf} \qsep \exists \widetilde{\vars}_\rmt, t_\rmf \in \timespan(\timedword) \qsep
			\transform_\mathrm{init}^\rmd(\TS) \land \shake_{0.1\second}(\timedword) \land \shake_{1\metre} \land \transform_\mathrm{word}^{\rmd,\tilde{\rmt}}(\sigma,t_\rmf) \land \lnot\transform_\rmf^{\tilde{\rmd}}(\Upsilon, \safe) \enspace,
\]
where $\Upsilon = (\Caridentifiers, 1,3, \tilde{x}_{\rml,\rmf}, \tilde{x}_{\rmr,\rmf}, \val)$.
Further, we point out that similar to Example~\ref{ex:transformation-potential-collision} we have $\lnot\transform_\rmf^{\tilde{\rmd}}(\Upsilon, \safe) = \transform_\rmf^{\tilde{\rmd}}(\Upsilon, \lnot\safe)$. 

We give an explanation of how our construction works for this example.
The perturbed variables $\tilde{t}_{\carD,2}, \tilde{t}_{\carE,2}$ from $\widetilde{vars}_\rmt$ represent perturbations of the time stamps $\tau_{\carD,1} = 1\second$ and $\tau_{\carE,1.1\second}$.
For the perturbed variables we choose $\assignment{\tilde{t}_{\carD,2}} = 1.1\second$ and $\assignment{\tilde{t}_{\carE,2}} = 1\second$.
This ensures that the order of the perturbed time stamps is switched.
The resulting perturbed timed word is
\[
	\timedword' = (\setReservation(\carE),1\second) \ (\removeReservation(\carD,3),1.1\second)\ (\removeReservation(\carE,2),6.1\second) \ (\wordend,6.1\second) \enspace.
\]
When applying $\timedword'$ to $\TS$ we get a transition sequence where car $\carD$ and car $\carE$ both simultaneously have two reservations for a duration of  $0.1\second$.

We choose to evaluate $\lnot \mathrm{safe}$ at $\assignment{t_\rmf} = 1\second$, i.e.\ after $\carE$ sets a new reservation and before $\carD$ withdraws it reservation, and discard all later changes. 
After $1\second$ car $\carD$ is at position $\assignment{\pos_{\carD,\rmf}} = 34\metre$ and car $\carE$ is at position $\assignment{\pos_{\carE,\rmf}} = 18\metre$ with a braking distance of $\assignment{\sensorfunction_{\carE,\rmf}} = 15\metre$.
We perturb $\pos_{\carD,\rmf}$ by $-1\metre$ and $\sensorfunction_{\carE,\rmf}$ by $+1\metre$.
The other variables are not perturbed.
We get $\assignment{\widetilde{\pos}_{\carD,\rmf}} = 33\metre$ and  $\assignment{\widetilde{\sensorfunction}_{\carE,\rmf}} = 16\metre$.

After moving the extension along with $\carE$ and not perturbing it, we evaluate $\transform_\rmf^{\tilde{\rmd}}(\Upsilon,\lnot \mathrm{safe})$ with 
	$\Upsilon = (\Caridentifiers, 1,3,\tilde{x}_{\rml,\rmf}\tilde{x}_{\rmr,\rmf},\val)$,
	$\assignment{\tilde{x}_{\rml,\rmf}} = 12\metre$, 
	$\assignment{\tilde{x}_{\rmr,\rmf}} = 102\metre$ on the perturbed variables $\widetilde{\vars}_{\rmd,\rmf}$. 
The formula $\lnot \safe$ states that there is a subview where the reservations of two different cars overlap.
As (after perturbation) car $\carD$ and car $\carE$ both have a reservation on lane $2$ and as the position of $\carD$ ($33\metre$) is within the space reserved by $\carE$ ($[18\metre,18\metre + 16\metre]$), the formula $\lnot \mathrm{safe}$ is satisfied.
Hence, $\timedword(M) \models_\mathrm{seq}^{\epsilon,\delta} \ltlglobally \mathrm{safe}$ does not hold. 
\end{example}

\section{Discussion}
Spatio-temporal robustness has been studied before for more abstract formalisms \cite{DM10,Que13}.
However, here the data for which we want to achieve robustness has a specific meaning, i.e.\ the underlying  model of MLSL is dedicated to modelling motorway traffic.
To this end, we study spatio-temporal robustness, taking the meaning of data into account.

%
In real-time systems we distinguish between time-driven and event-driven real-time systems \cite{Kop91}.
In MLSL we have two kinds of data values: the event-driven values $\clm$, $\res$ and $\acc$ and the time-driven values $\pos$, $\spd$ and $\sensorfunction$.
We study temporal robustness only for the event-driven values $\clm$ and $\res$.
For this we use the methodology from timed languages, where time stamps are perturbed \cite{GHJ97}.
Additionally, we study spatial robustness for the time-driven values $\pos$ and $\sensorfunction$ in a static `timeless' manner  at the level of traffic snapshots.
In \cite{FP09} such a `timeless' approach to spatial robustness has been done for Metric Temporal Logic.

%
%

One of the goals in the definition of MLSL was to reduce complexity of spatial reasoning by separating the spatial aspects from the car dynamics \cite{HLOR11}.
In this sense, the introduction of temporal robustness by perturbing time stamps seems well suited for MLSL, because we separate temporal robustness from spatial robustness, which simplifies reasoning.

A disadvantage of our approach is that at the linking of time-driven and event-driven values (here $\acc$ values, as they are event-driven and affect future evolution of time-driven values) we do not achieve temporal robustness, as it affects spatial robustness.

For our approach to temporal robustness we consider similarity of timed words.
A common definition to quantify similarity of timed words is defined in \cite{GHJ97}.
However, there the requirement is made that timed words have an infinite distance if they do not agree on the order of events.
In \cite{AFS04} the authors define a quantitative notion of (bi)similarity.
However, they define that the $i$th position in one sequence is compared to the $i$th position in another sequence, i.e.\ they do not consider that the order of events may not always be relevant.
Here the timed words originate from a distributed system, which makes it unreasonable to always consider the order of events as relevant.
Hence, we define an independence relation in the sense of \cite{Maz86} and in our quantification of similarity allow independent events to change their order.
To the best of our knowledge, a quantitative comparison of timed words under consideration of causality has not been used before.

On the side of efficiency we add the following observation:
If in the static MLSL formula $\phi$ all horizontal chop operators are below an odd number of negations, then the arithmetic formulas $\lnot \transform_\ltlglobally(\timedword, M, \phi)$ and $\lnot \transform_\ltlglobally^{\epsilon,\delta}(\timedword, M, \phi)$ only contain existentially quantified variables over the reals.
If we use an SMT solver \cite{BSST09} to check satisfiability of the formulas, we can interpret all variables as \emph{uninterpreted constants}, for which the solver tries to find a satisfying assignment.
This yields a significant speedup.

\section{Conclusion}

%

In this work we define a linear version of a dense-time globally operator for MLSL.
While there has been a temporal extension of MLSL \cite{LH15}, it is not suitable for monitoring, as it is a branching time temporal extension.
Our first main result is a transformation that takes an MLSL transition sequence $\timedword(M)$ and an MLSL formula $\phi$ to create a formula from the decidable first-order theory of real-closed fields \cite{Tar51}, such that the resulting formula is valid iff $\phi$ holds globally in $\timedword(M)$.

We then extend our transformation to accomodate for imprecise spatio-temporal data.
For this we defined a causality respecting notion of spatio-temporal similarity, which we base on timed words.
Our second main result is a transformation that additionally to the transition sequence $\timedword(M)$ and the static MLSL formula $\phi$ takes a maximal temporal error $\epsilon$ and a maximal spatial error $\delta$, such that the resulting formula is valid iff $\ltlglobally \phi$ holds $\epsilon$-$\delta$-robustly in $\timedword(M)$.
Again, the resulting formula is from the first-order theory of real-closed fields, and can algorithmically be checked for satisfiability.

Note that, while we consider only uni-directional traffic, our results easily extend to bi-directional traffic.
Speed and acceleration, the braking distance and the physical length of a car would then take negative values for cars going in the other direction and need to be updated accordingly when a car starts driving in the other direction.

In this work we define only a linear time globally operator for MLSL.
For future work we would like to define a fully fledged temporal extension of MLSL, where temporal operators  are basically taken from Metric Temporal Logic \cite{Koy90} and atoms are MLSL formulas.
It is desirable to extend our transformation to such an extended temporal version of MLSL.

We stated several claims in this work.
However, we did not provide proofs for them.
In future work proofs for our claims are certainly desirable.

Another line of research is to create temporal signals for MLSL formulas.
Such a temporal signal then represents for every instant in time if the MLSL formula currently is satisfied.
Then, we could use the significant work done for monitoring of Metric Temporal Logic \cite{FP09,FP06,DM10} and similar logics.


%% file: main.bbl
\begin{thebibliography}{10}
\providecommand{\bibitemdeclare}[2]{}
\providecommand{\surnamestart}{}
\providecommand{\surnameend}{}
\providecommand{\urlprefix}{Available at }
\providecommand{\url}[1]{\texttt{#1}}
\providecommand{\href}[2]{\texttt{#2}}
\providecommand{\urlalt}[2]{\href{#1}{#2}}
\providecommand{\doi}[1]{doi:\urlalt{http://dx.doi.org/#1}{#1}}
\providecommand{\bibinfo}[2]{#2}

\bibitemdeclare{inproceedings}{AFS04}
\bibitem{AFS04}
\bibinfo{author}{Luca \surnamestart de~Alfaro\surnameend},
  \bibinfo{author}{Marco \surnamestart Faella\surnameend} \&
  \bibinfo{author}{Mari{\"e}lle \surnamestart Stoelinga\surnameend}
  (\bibinfo{year}{2004}): \emph{\bibinfo{title}{Linear and {{Branching
  Metrics}} for {{Quantitative Transition Systems}}}}.
\newblock In \bibinfo{editor}{Josep \surnamestart D\'{\i}az\surnameend},
  \bibinfo{editor}{Juhani \surnamestart Karhum{\"a}ki\surnameend},
  \bibinfo{editor}{Arto \surnamestart Lepist{\"o}\surnameend} \&
  \bibinfo{editor}{Donald \surnamestart Sannella\surnameend}, editors: {\sl
  \bibinfo{booktitle}{{{ICALP}}}}, {\sl \bibinfo{series}{LNCS}}
  \bibinfo{volume}{3142}, \bibinfo{publisher}{{Springer}}, pp.
  \bibinfo{pages}{97--109}, \doi{10.1007/978-3-540-27836-8_11}.

\bibitemdeclare{article}{AD14}
\bibitem{AD14}
\bibinfo{author}{Matthias \surnamestart Althoff\surnameend} \&
  \bibinfo{author}{John~M \surnamestart Dolan\surnameend}
  (\bibinfo{year}{2014}): \emph{\bibinfo{title}{Online Verification of
  Automated Road Vehicles Using Reachability Analysis}}.
\newblock {\sl \bibinfo{journal}{IEEE Transactions on Robotics}}
  \bibinfo{volume}{30}(\bibinfo{number}{4}), pp. \bibinfo{pages}{903--918},
  \doi{10.1109/TRO.2014.2312453}.

\bibitemdeclare{article}{AD94}
\bibitem{AD94}
\bibinfo{author}{Rajeev \surnamestart Alur\surnameend} \&
  \bibinfo{author}{David~L. \surnamestart Dill\surnameend}
  (\bibinfo{year}{1994}): \emph{\bibinfo{title}{A {{Theory}} of {{Timed
  Automata}}}}.
\newblock {\sl \bibinfo{journal}{Theor. Comput. Sci.}}
  \bibinfo{volume}{126}(\bibinfo{number}{2}), pp. \bibinfo{pages}{183--235},
  \doi{10.1016/0304-3975(94)90010-8}.

\bibitemdeclare{incollection}{BSST09}
\bibitem{BSST09}
\bibinfo{author}{Clark~W. \surnamestart Barrett\surnameend},
  \bibinfo{author}{Roberto \surnamestart Sebastiani\surnameend},
  \bibinfo{author}{Sanjit~A. \surnamestart Seshia\surnameend} \&
  \bibinfo{author}{Cesare \surnamestart Tinelli\surnameend}
  (\bibinfo{year}{2009}): \emph{\bibinfo{title}{Satisfiability Modulo
  Theories}}.
\newblock In \bibinfo{editor}{Armin \surnamestart Biere\surnameend},
  \bibinfo{editor}{Marijn \surnamestart Heule\surnameend},
  \bibinfo{editor}{Hans \surnamestart van Maaren\surnameend} \&
  \bibinfo{editor}{Toby \surnamestart Walsh\surnameend}, editors: {\sl
  \bibinfo{booktitle}{Handbook of Satisfiability}}, {\sl
  \bibinfo{series}{Frontiers in Artificial Intelligence and Applications}}
  \bibinfo{volume}{185}, \bibinfo{publisher}{{IOS} Press}, pp.
  \bibinfo{pages}{825--885}, \doi{10.3233/978-1-58603-929-5-825}.

\bibitemdeclare{inproceedings}{DM10}
\bibitem{DM10}
\bibinfo{author}{Alexandre \surnamestart Donz{\'e}\surnameend} \&
  \bibinfo{author}{Oded \surnamestart Maler\surnameend} (\bibinfo{year}{2010}):
  \emph{\bibinfo{title}{Robust {{Satisfaction}} of {{Temporal Logic}} over
  {{Real}}-{{Valued Signals}}}}.
\newblock In \bibinfo{editor}{Krishnendu \surnamestart Chatterjee\surnameend}
  \& \bibinfo{editor}{Thomas~A. \surnamestart Henzinger\surnameend}, editors:
  {\sl \bibinfo{booktitle}{{{FORMATS}}}}, {\sl \bibinfo{series}{LNCS}}
  \bibinfo{volume}{6246}, \bibinfo{publisher}{{Springer}}, pp.
  \bibinfo{pages}{92--106}, \doi{10.1007/978-3-642-15297-9_9}.

\bibitemdeclare{inproceedings}{FP06}
\bibitem{FP06}
\bibinfo{author}{Georgios~E. \surnamestart Fainekos\surnameend} \&
  \bibinfo{author}{George~J. \surnamestart Pappas\surnameend}
  (\bibinfo{year}{2006}): \emph{\bibinfo{title}{Robustness of {{Temporal Logic
  Specifications}}}}.
\newblock In \bibinfo{editor}{Klaus \surnamestart Havelund\surnameend},
  \bibinfo{editor}{Manuel \surnamestart N{\'u}{\~n}ez\surnameend},
  \bibinfo{editor}{Grigore \surnamestart Ro\c{}su\surnameend} \&
  \bibinfo{editor}{Burkhart \surnamestart Wolff\surnameend}, editors: {\sl
  \bibinfo{booktitle}{{{FATES}}}}, \bibinfo{series}{LNCS},
  \bibinfo{publisher}{{Springer}}, pp. \bibinfo{pages}{178--192},
  \doi{10.1007/11940197_12}.

\bibitemdeclare{article}{FP09}
\bibitem{FP09}
\bibinfo{author}{Georgios~E. \surnamestart Fainekos\surnameend} \&
  \bibinfo{author}{George~J. \surnamestart Pappas\surnameend}
  (\bibinfo{year}{2009}): \emph{\bibinfo{title}{Robustness of Temporal Logic
  Specifications for Continuous-Time Signals}}.
\newblock {\sl \bibinfo{journal}{Theor. Comput. Sci.}}
  \bibinfo{volume}{410}(\bibinfo{number}{42}), pp. \bibinfo{pages}{4262--4291},
  \doi{10.1016/j.tcs.2009.06.021}.

\bibitemdeclare{inproceedings}{FH05}
\bibitem{FH05}
\bibinfo{author}{Martin \surnamestart Fr{\"a}nzle\surnameend} \&
  \bibinfo{author}{Michael~R. \surnamestart Hansen\surnameend}
  (\bibinfo{year}{2005}): \emph{\bibinfo{title}{A Robust Interpretation of
  Duration Calculus}}.
\newblock In \bibinfo{editor}{\surnamestart {Dang Van Hung}\surnameend} \&
  \bibinfo{editor}{\surnamestart {Martin Wirsing}\surnameend}, editors: {\sl
  \bibinfo{booktitle}{{{ICTAC}}}}, {\sl \bibinfo{series}{LNCS}}
  \bibinfo{volume}{3722}, \bibinfo{publisher}{{Springer}}, pp.
  \bibinfo{pages}{257--271}, \doi{10.1007/11560647_17}.

\bibitemdeclare{inproceedings}{FHO15}
\bibitem{FHO15}
\bibinfo{author}{Martin \surnamestart Fr{\"a}nzle\surnameend},
  \bibinfo{author}{Michael~R \surnamestart Hansen\surnameend} \&
  \bibinfo{author}{Heinrich \surnamestart Ody\surnameend}
  (\bibinfo{year}{2015}): \emph{\bibinfo{title}{No {{Need Knowing Numerous
  Neighbours}} - {{Towards}} a {{Realizable Interpretation}} of {{MLSL}}}}.
\newblock In \bibinfo{editor}{Roland \surnamestart Meyer\surnameend},
  \bibinfo{editor}{Andr{\'e} \surnamestart Platzer\surnameend} \&
  \bibinfo{editor}{Heike \surnamestart Wehrheim\surnameend}, editors: {\sl
  \bibinfo{booktitle}{Correct {{System Design}}}}, {\sl \bibinfo{series}{LNCS}}
  \bibinfo{volume}{9360}, \bibinfo{publisher}{{Springer}}, pp.
  \bibinfo{pages}{152--171}, \doi{10.1007/978-3-319-23506-6_11}.

\bibitemdeclare{inproceedings}{GHJ97}
\bibitem{GHJ97}
\bibinfo{author}{Vineet \surnamestart Gupta\surnameend},
  \bibinfo{author}{Thomas~A. \surnamestart Henzinger\surnameend} \&
  \bibinfo{author}{Radha \surnamestart Jagadeesan\surnameend}
  (\bibinfo{year}{1997}): \emph{\bibinfo{title}{Robust {{Timed Automata}}}}.
\newblock In \bibinfo{editor}{Oded \surnamestart Maler\surnameend}, editor:
  {\sl \bibinfo{booktitle}{Hybrid and {{Real}}-{{Time Systems}}}}, {\sl
  \bibinfo{series}{Lecture Notes in Computer Science}} \bibinfo{volume}{1201},
  \bibinfo{publisher}{{Springer}}, pp. \bibinfo{pages}{331--345},
  \doi{10.1007/BFb0014736}.

\bibitemdeclare{inproceedings}{HLOR11}
\bibitem{HLOR11}
\bibinfo{author}{M.~\surnamestart Hilscher\surnameend},
  \bibinfo{author}{S.~\surnamestart Linker\surnameend}, \bibinfo{author}{E.-R.
  \surnamestart Olderog\surnameend} \& \bibinfo{author}{A.~P. \surnamestart
  Ravn\surnameend} (\bibinfo{year}{2011}): \emph{\bibinfo{title}{An {{Abstract
  Model}} for {{Proving Safety}} of {{Multi}}-{{Lane Traffic Manoeuvres}}}}.
\newblock In \bibinfo{editor}{Shengchao \surnamestart Qin\surnameend} \&
  \bibinfo{editor}{Zongyan \surnamestart Qiu\surnameend}, editors: {\sl
  \bibinfo{booktitle}{{{ICFEM}}}}, {\sl \bibinfo{series}{LNCS}}
  \bibinfo{volume}{6991}, \bibinfo{publisher}{{Springer}}, pp.
  \bibinfo{pages}{404--419}, \doi{10.1007/978-3-642-24559-6_28}.

\bibitemdeclare{inproceedings}{HS16}
\bibitem{HS16}
\bibinfo{author}{Martin \surnamestart Hilscher\surnameend} \&
  \bibinfo{author}{Maike \surnamestart Schwammberger\surnameend}
  (\bibinfo{year}{2016}): \emph{\bibinfo{title}{An {{Abstract Model}} for
  {{Proving Safety}} of {{Autonomous Urban Traffic}}}}.
\newblock In \bibinfo{editor}{Augusto \surnamestart Sampaio\surnameend} \&
  \bibinfo{editor}{Farn \surnamestart Wang\surnameend}, editors: {\sl
  \bibinfo{booktitle}{{{ICTAC}}}}, {\sl \bibinfo{series}{LNCS}}
  \bibinfo{volume}{9965}, pp. \bibinfo{pages}{274--292},
  \doi{10.1007/978-3-319-46750-4_16}.

\bibitemdeclare{inproceedings}{Kop91}
\bibitem{Kop91}
\bibinfo{author}{Hermann \surnamestart Kopetz\surnameend}
  (\bibinfo{year}{1991}): \emph{\bibinfo{title}{Event-{{Triggered Versus
  Time}}-{{Triggered Real}}-{{Time Systems}}}}.
\newblock In \bibinfo{editor}{Arthur~I. \surnamestart Karshmer\surnameend} \&
  \bibinfo{editor}{J{\"u}rgen \surnamestart Nehmer\surnameend}, editors: {\sl
  \bibinfo{booktitle}{Operating {{Systems}} of the 90s and {{Beyond}}}}, {\sl
  \bibinfo{series}{LNCS}} \bibinfo{volume}{563},
  \bibinfo{publisher}{{Springer}}, pp. \bibinfo{pages}{87--101},
  \doi{10.1007/BFb0024530}.

\bibitemdeclare{article}{Koy90}
\bibitem{Koy90}
\bibinfo{author}{Ron \surnamestart Koymans\surnameend} (\bibinfo{year}{1990}):
  \emph{\bibinfo{title}{Specifying {{Real}}-{{Time Properties}} with {{Metric
  Temporal Logic}}}}.
\newblock {\sl \bibinfo{journal}{Real-Time Systems}}
  \bibinfo{volume}{2}(\bibinfo{number}{4}), pp. \bibinfo{pages}{255--299},
  \doi{10.1007/BF01995674}.

\bibitemdeclare{article}{LH15}
\bibitem{LH15}
\bibinfo{author}{Sven \surnamestart Linker\surnameend} \&
  \bibinfo{author}{Martin \surnamestart Hilscher\surnameend}
  (\bibinfo{year}{2015}): \emph{\bibinfo{title}{Proof {{Theory}} of a
  {{Multi}}-{{Lane Spatial Logic}}}}.
\newblock {\sl \bibinfo{journal}{Logical Methods in Computer Science}}
  \bibinfo{volume}{11}(\bibinfo{number}{3}), \doi{10.2168/LMCS-11(3:4)2015}.

\bibitemdeclare{inproceedings}{Maz86}
\bibitem{Maz86}
\bibinfo{author}{Antoni~W. \surnamestart Mazurkiewicz\surnameend}
  (\bibinfo{year}{1986}): \emph{\bibinfo{title}{Trace {{Theory}}}}.
\newblock In \bibinfo{editor}{Wilfried \surnamestart Brauer\surnameend},
  \bibinfo{editor}{Wolfgang \surnamestart Reisig\surnameend} \&
  \bibinfo{editor}{Grzegorz \surnamestart Rozenberg\surnameend}, editors: {\sl
  \bibinfo{booktitle}{Advances in {{Petri Nets}}}}, {\sl
  \bibinfo{series}{LNCS}} \bibinfo{volume}{255},
  \bibinfo{publisher}{{Springer}}, pp. \bibinfo{pages}{279--324},
  \doi{10.1007/3-540-17906-2_30}.

\bibitemdeclare{inproceedings}{Ody15}
\bibitem{Ody15}
\bibinfo{author}{Heinrich \surnamestart Ody\surnameend} (\bibinfo{year}{2015}):
  \emph{\bibinfo{title}{Undecidability {{Results}} for {{Multi}}-{{Lane Spatial
  Logic}}}}.
\newblock In \bibinfo{editor}{Martin \surnamestart Leucker\surnameend},
  \bibinfo{editor}{Camilo \surnamestart Rueda\surnameend} \&
  \bibinfo{editor}{Frank~D. \surnamestart Valencia\surnameend}, editors: {\sl
  \bibinfo{booktitle}{{{ICTAC}}}}, {\sl \bibinfo{series}{LNCS}}
  \bibinfo{volume}{9399}, \bibinfo{publisher}{{Springer}}, pp.
  \bibinfo{pages}{404--421}, \doi{10.1007/978-3-319-25150-9_24}.

\bibitemdeclare{phdthesis}{Que13}
\bibitem{Que13}
\bibinfo{author}{Jan-David \surnamestart Quesel\surnameend}
  (\bibinfo{year}{2013}): \emph{\bibinfo{title}{Similarity, {{Logic}}, and
  {{Games}} - {{Bridging Modeling Layers}} of {{Hybrid Systems}}}}.
\newblock Ph.D. thesis, \bibinfo{school}{University of Oldenburg}.

\bibitemdeclare{inproceedings}{RA15}
\bibitem{RA15}
\bibinfo{author}{Albert \surnamestart Rizaldi\surnameend} \&
  \bibinfo{author}{Matthias \surnamestart Althoff\surnameend}
  (\bibinfo{year}{2015}): \emph{\bibinfo{title}{Formalising Traffic Rules for
  Accountability of Autonomous Vehicles}}.
\newblock In: {\sl \bibinfo{booktitle}{{{ITSC}}}}, \bibinfo{publisher}{{IEEE}},
  pp. \bibinfo{pages}{1658--1665}, \doi{10.1109/ITSC.2015.269}.

\bibitemdeclare{misc}{Sch17}
\bibitem{Sch17}
\bibinfo{author}{Maike \surnamestart Schwammberger\surnameend}
  (\bibinfo{year}{2017}): \emph{\bibinfo{title}{Imperfect Knowledge in
  Autonomous Urban Traffic Manoeuvres}}.
\newblock \bibinfo{note}{FVAV}.

\bibitemdeclare{book}{Tar51}
\bibitem{Tar51}
\bibinfo{author}{Alfred \surnamestart Tarski\surnameend}
  (\bibinfo{year}{1951}): \emph{\bibinfo{title}{A Decision Method for
  Elementary Algebra and Geometry}}.
\newblock \bibinfo{publisher}{{University of California Press}}.

\end{thebibliography}
